\documentclass[12pt]{article}
\usepackage{cite,graphicx,epsfig}

\newcommand{\ket}{\left| I \right>}

\newcommand{\ba}{\begin{array}}
\newcommand{\ea}{\end{array}}
\newcommand{\be}{\begin{equation}}
\newcommand{\ee}{\end{equation}}
\newcommand{\bea}{\begin{eqnarray}}
\newcommand{\eea}{\end{eqnarray}}

\def\IB{\relax\hbox{$\inbar\kern-.3em{\rm B}$}}
\def\IC{\relax\hbox{$\inbar\kern-.3em{\rm C}$}}
\def\ID{\relax\hbox{$\inbar\kern-.3em{\rm D}$}}
\def\IE{\relax\hbox{$\inbar\kern-.3em{\rm E}$}}
\def\IF{\relax\hbox{$\inbar\kern-.3em{\rm F}$}}
\def\IG{\relax\hbox{$\inbar\kern-.3em{\rm G}$}}
\def\IGa{\relax\hbox{${\rm I}\kern-.18em\Gamma$}}
\def\IH{\relax{\rm I\kern-.18em H}}
\def\IK{\relax{\rm I\kern-.18em K}}
\def\IL{\relax{\rm I\kern-.18em L}}
\def\IP{\relax{\rm I\kern-.18em P}}
\def\IR{\relax{\rm I\kern-.18em R}}
\def\IZ{\relax{\rm Z\kern-.5em Z}}

\def\half{\frac{1}{2}}

\def\f{\frac}

\def\TL{Temperley-Lieb }

\begin{document}

\begin{titlepage}

\begin{flushright}

\end{flushright}

\vskip 2 cm

\begin{center}
{\LARGE One-boundary \TL algebras in the XXZ and loop models}
\vskip 1 cm

{\large A. Nichols\footnote{nichols@th.physik.uni-bonn.de},
  V. Rittenberg\footnote{vladimir@th.physik.uni-bonn.de}, and J. de
  Gier\footnote{degier@ms.unimelb.edu.au} }

\begin{center}
{\em $^{1,2}$Physikalisches Institut der Universit\"at Bonn, \\
 Nussallee 12, 53115 Bonn, Germany.}

{\em $^3$Department of Mathematics and Statistics, \\
University of Melbourne, Parkville,\\ Victoria VIC 3010, Australia.}

\end{center}

\vskip 1 cm


\begin{abstract}
We give an exact spectral equivalence between the quantum group
invariant XXZ chain with arbitrary left boundary term and the same XXZ
chain with purely diagonal boundary terms.

This equivalence, and a further one with a link pattern Hamiltonian, can be
understood as arising from different representations of the one-boundary
Temperley-Lieb algebra. For a system of size $L$ these representations 
are all of dimension $2^L$ and,
for generic points of the algebra, equivalent.  However at exceptional
points they can possess different indecomposable structures.

We study the centralizer of the one-boundary \TL algebra in the
`non-diagonal' spin-$\half$ representation and find its eigenvalues and
eigenvectors. In the exceptional cases the centralizer becomes
indecomposable. We show how to get a truncated space of `good' states.
The indecomposable part of the centralizer leads to degeneracies in the
three mentioned Hamiltonians.
\end{abstract}

\end{center}

\end{titlepage}

\newpage
\section{Introduction}
\renewcommand{\theequation}{\arabic{section}.\arabic{equation}}
\setcounter{equation}{0}  
The one-dimensional anisotropic spin-$\half$ Heisenberg model (the
XXZ quantum chain) is not only a paradigm for integrable systems but is also an
interesting model for describing experimental data \cite{Essler}. Recently it was
shown that for the special value of the anisotropy parameter
$\Delta=-\half$ the same Hamiltonian gives the time evolution of fluctuating
interfaces, known as the raise and peel models \cite{RaiseandPeel,PyatovHexagon}. Moreover in this case the ground-state
wavefunctions for various boundary conditions have remarkable combinatorial
properties \cite{Razumov:2000ei,JanReview}.

In the present paper we will present some new properties of the XXZ chain
with diagonal and non-diagonal boundary conditions. In particular we shall
give a spectral equivalence between the XXZ chain with different types of
boundary conditions. 

In the diagonal case, after the pioneering work of Alcaraz et al. \cite{Alcaraz:1987uk,Alcaraz:1988zr} on the
Bethe ansatz, many properties are known. In particular, for a special choice
of the boundary terms, the quantum chain has the
quantum symmetry $U_q(SU(2))$ \cite{Pasquier:1989kd}. For this case an alternative understanding of the properties of the chain
can be obtained from an algebraic point of view
as the terms appearing in the Hamiltonian are the
generators of the Temperley-Lieb (TL) algebra \cite{TemperleyLieb,MartinBook}. In this way, for
example, it was shown that the spectra of the Potts models were contained
in those of the XXZ chain \cite{Alcaraz:1987ni,Alcaraz:1988vi}. Away from the $U_q(SU(2))$ boundary
conditions, some degeneracies were observed numerically
\cite{Grimm:1990dg,Grimm:1990gg} but no explanation was found. 

The case of two general non-diagonal boundaries has received a lot of
attention more recently. Although is well known to be integrable \cite{deVega:1992zd} there is
no obvious Bethe reference state. At the decoupling point ($\Delta=0$) the
spectrum and wavefunctions have been obtained \cite{BilsteinI}. Away from this point, in
the case in which the parameters satisfy an additional constraint, the Bethe
ansatz equations were obtained using two different approaches. In the first
approach the Bethe ansatz was constructed directly using the quantum chain \cite{ChineseGuys,Nepomechie:2002xy,Nepomechie:2003vv,Nepomechie:2003ez}. In
a second, completely different approach, an `equivalent' Hamiltonian was
written in the vector space of link patterns \cite{deGier:2003iu}. In 
the two approaches one begins with a
Hamiltonian written in terms of generators of the two-boundary Temperley-Lieb
algebra (2BTL) \cite{JanReview} and uses two different representations of the algebra -
one acting in a spin basis and the other acting in a link pattern basis. 
As we shall explain one expects that although two Hamiltonians may have 
the same spectrum they may possess different Jordan cell structures making 
the number of eigenfunctions and the physics very different 
(\ref{sec:Jordan}). These aspects of the problem were not studied in 
\cite{ChineseGuys,Nepomechie:2002xy,Nepomechie:2003vv,Nepomechie:2003ez,deGier:2003iu}.

In the presentation of our work, we start in section $2$ by introducing the
one-boundary Temperley-Lieb algebra (1BTL), also known as the blob algebra, which is well known in the mathematical literature \cite{Martin:1992td,Martin:1993jk,MartinWoodcockI,MartinWoodcockII}. The 1BTL algebra depends on two parameters, one for the
bulk generators and one for the boundary generator. In the cases we shall call
`critical' (as coined in \cite{Martin:1992td,Martin:1993jk,MartinWoodcockI,MartinWoodcockII}) when a simple relation
between these two parameters is satisfied, the algebra becomes
non-semisimple, and has representations which are reducible but
indecomposable.

We start with the `master' Hamiltonian $H^M$ which is composed of 
$L-1$ bulk generators and one
boundary generator chosen by convention to be at the left end of the 1BTL. This
Hamiltonian has one explicit parameter (the coefficient of the boundary
generator) in addition to the two parameters of the 1BTL. We first use the
$L$-site `non-diagonal' representation of the 1BTL defined in the $2^L$ dimensional spin
basis. The one-boundary Hamiltonian is obtained by adding a general $2 \times 2$ matrix to the
end of the quantum group invariant Hamiltonian. It should be remembered that
the quantum group invariant Hamiltonian already has very particular diagonal
terms at either end. The one-boundary Hamiltonian $H^{nd}$ depends explicitly on three
parameters and has no local conserved charge. We show that the spectrum,
including degeneracies, of this Hamiltonian is identical with that of another
XXZ Hamiltonian $H^d$ defined in the same spin basis with general diagonal
boundary terms. We consider this observation as one of the most important
results of this paper. The relation between the spectra of the two
Hamiltonians using the Bethe ansatz is given in \ref{sec:Betheansatz}. We show, for the case of $2$ generators, that the 1BTL has also possesses a `diagonal' representation in the spin basis (for more on this topic see \ref{sec:DiagonalRepn}).

In Section 3 we introduce the representation of 1BTL algebra acting on
the space of link patterns (there are actually $2^L$ of them!) and obtain the Hamiltonian $H^{lp}$. The link
patterns have a diagrammatic charge with the same spectrum as $S^z$ of the quantum chain. However it is important to stress that this diagrammatic charge is not a conserved quantum number except in an ideal of the 1BTL corresponding to the charge zero sector. 

The Hamiltonian $H^{lp}$ has a lower block triangular structure. In order to
obtain its spectrum, one can disregard the off-diagonal blocks and obtain a `fake' Hamiltonian which commutes with the diagrammatic charge \cite{deGier:2003iu}. Although this Hamiltonian $H^{lp}$ has the same
spectrum as $H^{nd}$ and $H^d$ (see \ref{sec:Betheansatz}), the indecomposable
structures which can occur are different. Some simple examples are given to
illustrate this point. In \ref{sec:Similarities} we show in the case of $2$
sites, the similarity transformations which relate the three Hamiltonians away
from the `critical' cases. The
three representations of the 1BTL algebra, namely those in $H^{nd}$, $H^d$ and $H^{lp}$, are all of dimension $2^L$ and the issue of their
faithfulness is discussed in \ref{sec:Faithfulness}.

The problem of integrable quantum field theory on the half-line has led to the
discovery of `boundary quantum groups' \cite{Delius:2001qh,Delius:2002mv} generalizing earlier results at free fermion point \cite{Mezincescu:1997nw}. Based on these results Doikou \cite{Doikou:2004km} has found a centralizer of the 1BTL in the spin-$\half$
representation for the finite chain. This operator is not related to the
integrability of the model since it commutes with each member of the 1BTL algebra. In Section 4 we study
the eigenvalues and the eigenfunctions of this centralizer. For generic values
of the parameters of the 1BTL it is completely diagonalizable and its spectrum and degeneracies are related to
that of an $S^z$-type charge. However, for the cases in which the 1BTL algebra
is `critical', this centralizer is not fully diagonalizable. This explains the
appearance (see \ref{sec:Jordan}) of degeneracies in the spectrum of $H^{nd}$ and, via the spectral equivalence, also in $H^d$ and $H^{lp}$. Moreover the eigenfunctions of the centralizer allow a construction of the `good' representations of the 1BTL algebra. This
construction is based on truncated Bratelli diagrams and the results are
closely related to previous work by Martin et al. \cite{Martin:1992td,Martin:1993jk,MartinWoodcockI,MartinWoodcockII}.

In a similar way to the TL algebra, the 1BTL algebra also has Potts
representations which correspond to Potts models with boundary terms. These
models will be considered elsewhere \cite{WorkInProgress}. Conclusions and open questions will be presented in Section $5$.
\section{The one-boundary \TL algebra and XXZ quantum chains}
\setcounter{equation}{0}  
We would like to remind the reader of some known facts about the connections
between the Temperley-Lieb algebra and the XXZ quantum chain \cite{MartinBook}.
 The Temperley-Lieb algebra (TL) is an associative algebra with the generators
 $e_i$ $(i=1,\cdots,L-1)$ obeying the relations:
\bea \label{eqn:TL}
e_i e_{i\pm 1} e_i&=&e_i \nonumber \\
e_i e_j& =& e_j e_i \quad |i-j|>1\\    
e_i^2&=&(q+q^{-1})~e_i\nonumber
\eea
%
This has a representation in terms of Pauli matrices:
\bea \label{eqn:TLgenerators}
e_i= \half \left\{ \sigma^x_i \sigma^x_{i+1} + \sigma^y_i \sigma^y_{i+1} - \cos \gamma \sigma^z_i \sigma^z_{i+1}  + \cos \gamma + i \sin \gamma \left(\sigma^z_i - \sigma^z_{i+1} \right) \right\}
\eea
with $q=e^{i\gamma}$. Using these generators we can define the quantum group
invariant Hamiltonian $H^{qg}$ of a ferromagnetic quantum chain:
\bea \label{eqn:qgHamiltonian}
H^{qg}&=&-\sum_{i=1}^{L-1} e_i \nonumber \\
&=& -\half \left\{ \sum_{i=1}^{L-1} \left( \sigma^x_i \sigma^x_{i+1} + \sigma^y_i \sigma^y_{i+1} - \cos \gamma \sigma^z_i \sigma^z_{i+1} + \cos \gamma \right) + i \sin \gamma \left(\sigma^z_1 - \sigma^z_L \right) \right\}~~~
\eea
This integrable Hamiltonian, defined on the $2^L$ dimensional vector space, has an anisotropy
parameter $\Delta = -\cos \gamma$ and is $U_q(SU(2))$ symmetric \cite{Pasquier:1989kd}. As is well known \cite{Pasquier:1989kd}, if $q$ is a root of unity the
quantum group has `good' (irreducible) representations as well as `bad'
(indecomposable) representations. As a consequence of the existence of `bad'
representations (see \ref{sec:Jordan}), the spectrum of $H^{qg}$ when $q$ is a root
of unity has higher degeneracies than occur for generic $q$. In
this case it also has indecomposable structure (the number of eigenfunctions is smaller
than $2^L$). 

The one-boundary Temperley-Lieb algebra (1BTL) is obtained \cite{Martin:1992td,Martin:1993jk,MartinWoodcockI,MartinWoodcockII} by adding
a new generator $e_0$ to the TL algebra. It has the following additional relations:
\bea \label{eqn:BoundaryTL}
e_1 e_0 e_1&=&e_1 \nonumber \\
e_0^2&=&\f{\sin \omega}{\sin(\omega+\gamma)} e_0 \\
e_0 e_i&=& e_i e_0 \quad i>1 \nonumber
\eea
Notice that in addition to the bulk parameter $\gamma$ the 1BTL algebra has a
second parameter $\omega$ which is only defined up to a multiple of
$\pi$. These can both be complex in general but for convenience, we will assume
that they are real in order to use trigonometric functions. 

It was observed by Martin et al. \cite{Martin:1992td,Martin:1993jk,MartinWoodcockI,MartinWoodcockII} that if:
\bea \label{eqn:boundaryTLcritical}
\omega = k \gamma + \pi {\bf Z}
\eea
with $k$ integer, then the 1BTL algebra becomes non-semisimple and possesses
indecomposable representations. We shall keep the rather unusual name
`critical' introduced in \cite{MartinWoodcockI,MartinWoodcockII}. The case $\omega = -\gamma$, which is also
`critical', requires a rescaling of the generator $e_0$ in
(\ref{eqn:BoundaryTL}). One should note that the
indecomposability is controlled by $\omega$ and it can be `critical' (called
simple critical in \cite{Martin:1992td,Martin:1993jk,MartinWoodcockI,MartinWoodcockII}) even when $\gamma$ is generic i.e. $q=e^{i \gamma}$ generic,
as long as the relation (\ref{eqn:boundaryTLcritical}) is satisfied.

Using the generators of the 1BTL algebra, we consider the `Master' Hamiltonian:
\bea \label{eqn:TLlb}
H^{M}&=&-a e_0-\sum_{i=1}^{L-1} e_i
\eea
where $a$ is an arbitrary parameter.

As we are going to see, due to the existence of several different
representations of the 1BTL algebra the properties of this Hamiltonian in different $2^L$
dimensional vector spaces are different. The spectra are the same but the
Jordan cell structures are different.

The first representation of this algebra, which we shall call `non-diagonal', is obtained taking the bulk $e_i$
 with $1 \le i \le L-1$ as in (\ref{eqn:TLgenerators}) and:
\bea
e_0&=&-\half \f{1}{\sin(\omega+\gamma)}\left( i \cos \omega \sigma_1^z + \cos
  \phi \sigma_1^x + \sin \phi \sigma_1^y - \sin \omega \right)
\eea
The angle $\phi$ is irrelevant as it can be changed by a rotation of $\sigma^x_1$
and $\sigma^y_1$ preserving the bulk generators (\ref{eqn:TLgenerators}). In
this paper we shall put $\phi=0$. Then we have:
\bea \label{eqn:e0}
e_0&=&-\half \f{1}{\sin(\omega+\gamma)}\left( i \cos \omega \sigma_1^z +  \sigma_1^x  - \sin \omega \right) \nonumber \\
&=&-\half \f{1}{\sin(\omega+\gamma)}\left(
\begin{array}{cc}
i e^{i \omega} & 1  \\
1 & -i e^{-i \omega}
\end{array}
\right) \otimes {\bf 1} \otimes \cdots \otimes {\bf 1}
\eea
In this way we obtain:
\bea \label{eqn:Hnd}
H^{nd}&=& \f{\sin \gamma}{\cos \omega +\cos \delta}\left( i \cos \omega \sigma_1^z + \sigma_1^x  - \sin \omega \right)\nonumber\\
&&-\half \left\{ \sum_{i=1}^{L-1} \left( \sigma^x_i \sigma^x_{i+1} +
    \sigma^y_i \sigma^y_{i+1} - \cos \gamma \sigma^z_i \sigma^z_{i+1} + \cos
    \gamma \right) + i \sin \gamma \left(\sigma^z_1 - \sigma^z_L \right)
\right\}~~~
\eea
where we have used the following convenient parameterization for $a$:
\bea \label{eqn:Definitionofa}
a= \f{2 \sin \gamma \sin(\omega+\gamma)}{\cos \omega + \cos \delta}
\eea
Notice that $H^{nd}$ is dependent on three parameters: $\gamma$ and $\omega$,
related to the algebra and $\delta$ related to the constant $a$ in the
Hamiltonian (\ref{eqn:TLlb}). We shall see that their roles in the physical
properties of $H^{nd}$ are different.
 
It is interesting to observe that on the first site, $H^{nd}$ contains the most
general boundary term. This is not the case for the last site where $H^{nd}$ has
the same boundary term as $H^{qg}$ c.f. (\ref{eqn:qgHamiltonian}) and
(\ref{eqn:Hnd}). One can see that there is no local charge which commutes with $H^{nd}$.

One of the main results of this paper is that for any values
of the three parameters, the spectrum of $H^{nd}$ exactly coincides with the
spectrum of the Hamiltonian $H^d$ with diagonal boundary terms only:
\bea \label{eqn:Hd}
H^{d}&=&-\half \left\{ \sum_{i=1}^{L-1} \left( \sigma^x_i \sigma^x_{i+1} + \sigma^y_i \sigma^y_{i+1} - \cos \gamma \sigma^z_i \sigma^z_{i+1} + \cos \gamma \right) \right.\nonumber\\
&&\left.+\sin \gamma \left[\tan \left(\f{\omega+\delta}{2}\right) \sigma_1^z +
    \tan \left(\f{\omega-\delta}{2} \right)\sigma_L^z  +\f{2 \sin \omega}{\cos
      \omega +\cos \delta} \right] \right\}~~~
\eea
We postpone for a moment the proof of this statement. Unlike $H^{nd}$, the diagonal chain has the obvious local charge:
\bea \label{eqn:Sz}
S^z =\half \sum_{i=1}^L \sigma_i^z
\eea
The fact that the spectra are identical does not necessarily mean that in the
whole parameter space, one can construct a similarity transformation relating
these two Hamiltonians. In order to illustrate this point, in \ref{sec:Similarities} we
derive the similarity transformation which relates $H^{nd}$ to $H^{d}$ for two sites. The similarity transformation
exists everywhere except for $\omega=-\gamma,0,\gamma$. For a larger number of
sites there are many more possible values of $\omega$ where a similarity
transformation would break down. At these points the two
Hamiltonians have repeated eigenvalues and different Jordan cell
structures. In many examples, by exact diagonalization, we found that $H^{nd}$ always has Jordan
cell structures for these cases and $H^d$, being Hermitian at least for real
values of the parameters $\gamma, \omega$ and $\delta$, does not.

We have to keep in mind that whereas $H^{nd}$ was obtained using (\ref{eqn:TLlb}) and a
representation of the 1BTL algebra, $H^d$ came from nowhere. In \ref{sec:DiagonalRepn},
it is shown in the case of 2 sites, that the 1BTL has a different
representation in the space of $4 \times 4$ matrices which, using
(\ref{eqn:TLlb}), gives precisely $H^d$ for two sites. We also discuss the generalization of this observation.

In \ref{sec:Betheansatz} we prove using the Bethe ansatz that the spectra of
$H^{nd}$ and $H^{d}$ coincide. The proof has a subtle point. 
Contrary to the case of periodic boundary conditions
\cite{Baxter:2001sx}, the derivation of the Bethe Ansatz equations for $H^d$ seems
to be valid only on `one side of the equator'. If one takes the eigenstate
with $S^z = \f{L}{2}$ as a reference state, the Bethe ansatz only
gives the correct eigenvalues for $S^z \ge 0$. The remaining eigenvalues
for $S^z < 0$ are obtained by employing again the Bethe ansatz but
taking as a reference state the eigenstate with $S^z = -\f{L}{2}$. 

The Jordan cell structures of $H^{nd}$ and their connection with the
`critical' algebras are going to be derived in Section 4. The Jordan
cells of $H^{nd}$ `induce' degeneracies in $H^d$.

In the next Section we are going to present another representation of
the 1BTL algebra in a $2^L$ vector space and the corresponding Hamiltonian.
This Hamiltonian has the same spectrum as both $H^{nd}$ and $H^d$ but the
indecomposable structures can be different.
\section{The link pattern representation of the one-boundary \TL algebra}
\setcounter{equation}{0}  
\label{sec:Loop}
A different representation of the 1BTL algebra in another $2^L$ dimensional
vector space is obtained if we use link patterns. This representation is
related to loop models \cite{JanReview} and has been used in stochastic models \cite{PyatovHexagon}
and combinatorics \cite{JanReview}.

We start by giving the standard graphical representation of the 1 BTL algebra:
\bea
e_i \quad &=&\quad
\begin{picture}(130,20)
\put(0,0){\epsfxsize=130pt\epsfbox{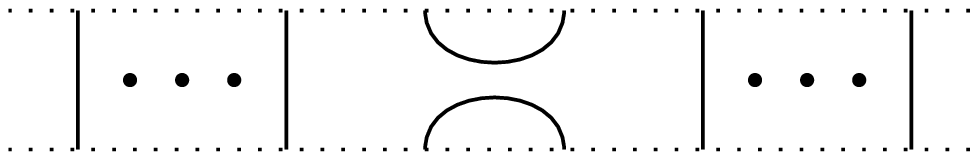}}
\put(51,-10){{\small i}}
\put(67,-10){{\small i+1}}
\end{picture} \\
\nonumber \\
e_0 \quad &=&\quad
\begin{picture}(240,20)
\put(0,0){\epsfxsize=70pt\epsfbox{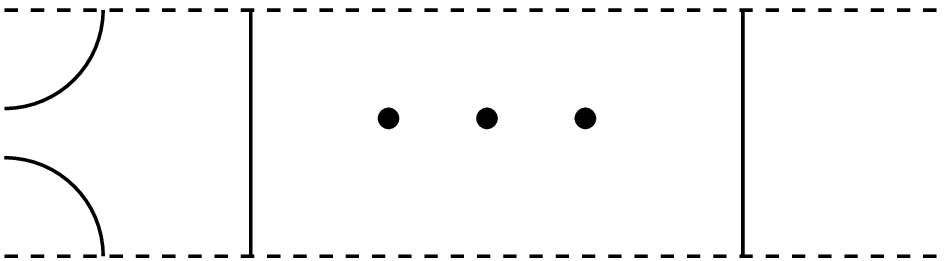}}
\put(40,-10){{\small 1}}
\end{picture}
\label{eq:f1}
\eea
Multiplication of two words in the algebra corresponds to putting one
word below the other and merging the loops lines. For example, the
relations $e_i^2=2 \cos \gamma e_i$, $e_0^2=\f{\sin \omega}{\sin(\omega+\gamma)} e_0$ and $e_ie_{i+1}e_i=e_i$ graphically read:
\bea
\begin{picture}(120,40)
\put(0,0){\epsfxsize=40pt\epsfbox{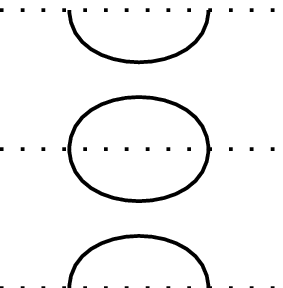}}
\put(50,18){$=2 \cos \gamma$}
\put(100,10){\epsfxsize=40pt\epsfbox{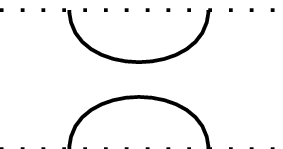}}
\end{picture}
\label{eq:e^2}
\eea
\vskip2mm
\bea
\begin{picture}(240,20)
\put(0,0){\epsfxsize=70pt\epsfbox{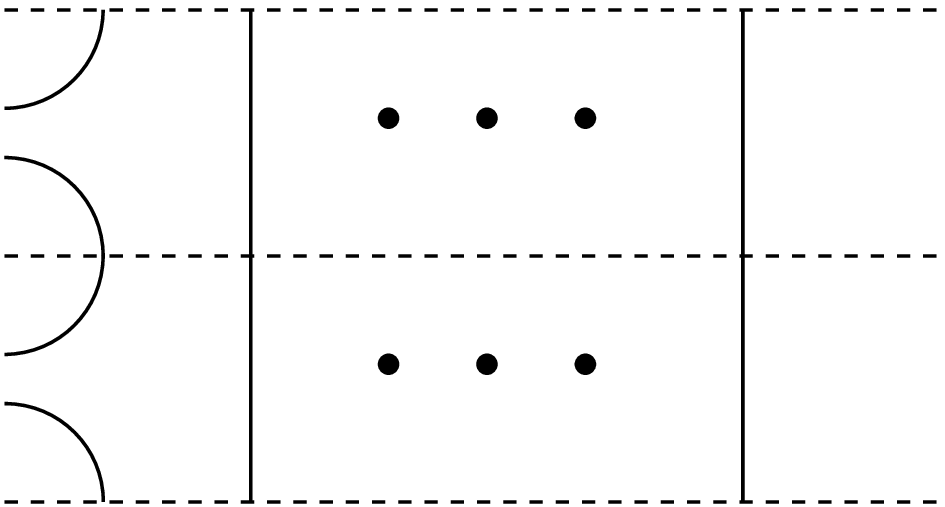}}
\put(80,18){$=\f{\sin \omega}{\sin(\omega+\gamma)}$} 
\put(140,10){\epsfxsize=70pt\epsfbox{e0.eps}}
\end{picture}
\label{eqn:e0^2}
\eea
\vskip2mm
\bea
\begin{picture}(150,60)
\put(0,0){\epsfxsize=60pt\epsfbox{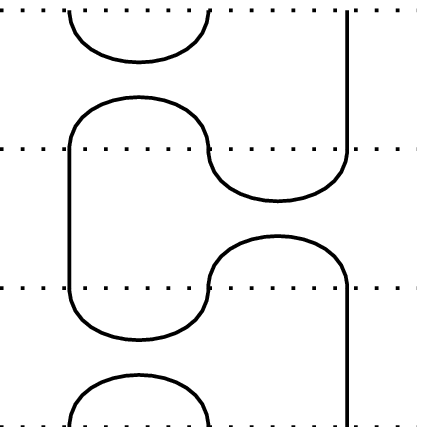}}
\put(70,28){$=$}
\put(90,20){\epsfxsize=60pt\epsfbox{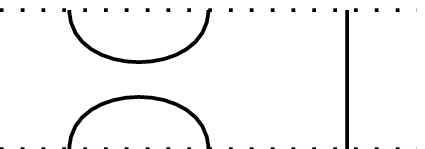}}
\end{picture}
\eea
\vskip 6mm
The link pattern representation corresponds to considering an ideal of
the 1BTL. We consider the state $\ket$
 by taking the graph corresponding to the unit element $\bf{1}$ of the 1BTL algebra. It
 has all $L$ sites unconnected:
\vskip 2mm
\bea
\ket  \quad = \quad \begin{picture}(140,10)
\put(0,0){\epsfxsize=70pt\epsfbox{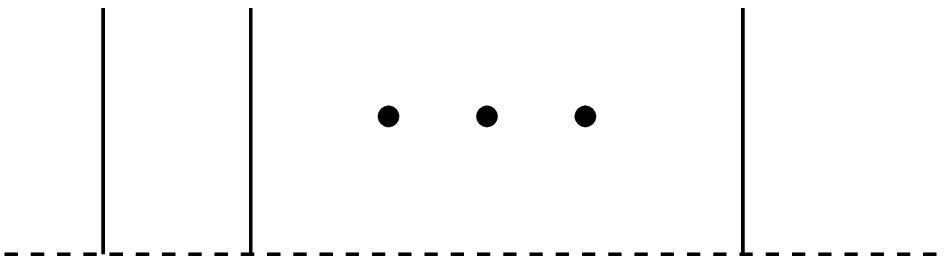}}
\end{picture}
\eea
We then act with the algebra and finally keep only the bottom half of the picture:
\bea
{\bf 1} \ket  \quad = \quad \begin{picture}(140,10)
\put(0,0){\epsfxsize=70pt\epsfbox{Ideal.eps}}
\end{picture}
\eea
\bea
e_i \ket \quad = \quad
\begin{picture}(140,10)
\put(0,0){\epsfxsize=140pt\epsfbox{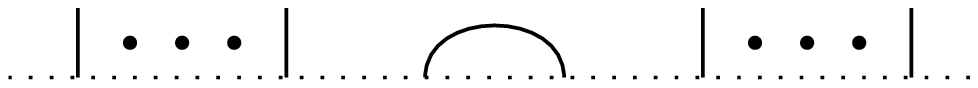}}
\put(56,-10){{\small i}}
\put(72,-10){{\small i+1}}
\end{picture}
\label{eq:eiT} \\
\nonumber \\
e_0 \ket \quad= \quad 
\begin{picture}(140,10)
\put(0,0){\epsfxsize=70pt\epsfbox{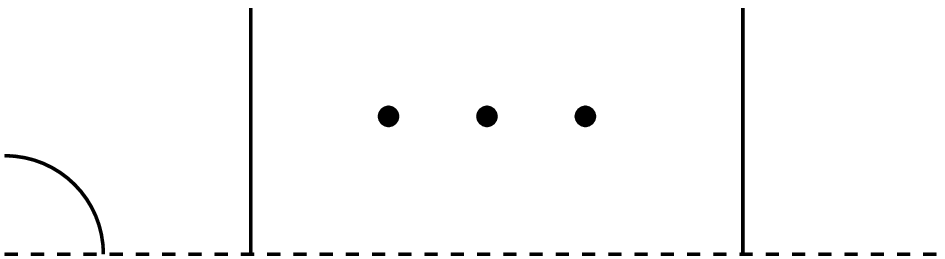}}
\end{picture}
\eea
\vskip 6mm
Other examples are:
\bea
e_{i+1}e_i \ket \quad &=&\quad 
\begin{picture}(160,30)
\put(0,0){\epsfxsize=160pt\epsfbox{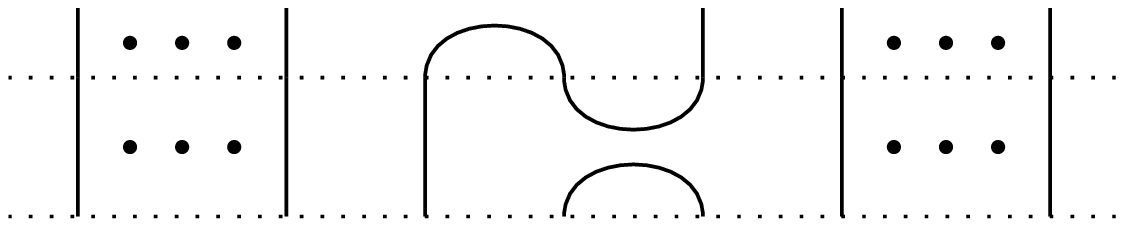}}
\put(56,-10){{\small i}}
\put(72,-10){{\small i+1}}
\put(92,-10){{\small i+2}}
\end{picture} \nonumber\\[14pt]
\nonumber \\
&=& \quad
\begin{picture}(160,10)
\put(0,0){\epsfxsize=160pt\epsfbox{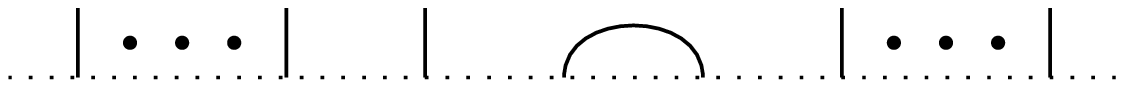}}
\put(56,-10){{\small i}}
\put(72,-10){{\small i+1}}
\put(95,-10){{\small i+2}}
\end{picture}
\label{eq:eip1T}
\eea
\vskip 6mm
and:
\bea
e_{i+1}e_{i+2}e_i \ket \quad = \quad
\begin{picture}(170,15)
\put(0,0){\epsfxsize=170pt\epsfbox{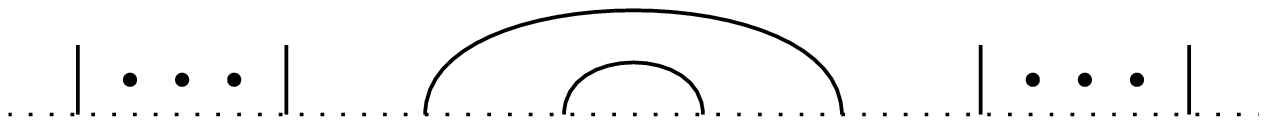}}
\put(52,-10){{\small i}}
\put(106,-10){{\small i+3}}
\end{picture}\;.
\label{eq:eiTcomp}
\eea
\vskip 6mm
Instead of the graphical representation we use a more convenient typographical notation. If a site $i$ is connected to the left or right we write ``$)$'' and ``$($'' respectively. If it is unconnected then we write ``$|$''. Thus the picture (\ref{eq:eiTcomp}) is written ``$|\cdots|(())| \cdots |$''. As we are dealing here with the 1BTL we \emph{cannot} have links to the right boundary. 

It is natural to introduce a diagrammatic charge, $C$, which counts the number of
sites which contain links connected to another site or to the boundary. The link patterns form a vector space of dimension $2^L$ and the
diagrammatic charge $C$ splits it into subspaces having the dimension of the
binomial coefficients. More specifically for a system of size $L$ we have
$C=0,1,\cdots,L$ with:
\begin{itemize}
\item{$C$ even}
\bea \label{eqn:Diagchargespaces}
{\rm dimension}=\left(
\begin{array}{c}
L \\
\f{L}{2} - C 
\end{array}
\right)
\eea
\item{$C$ odd}
\bea
{\rm dimension}=\left(
\begin{array}{c}
L \\
-\f{L}{2} + C - 1  
\end{array}
\right)
\eea
\end{itemize}
This
is illustrated for the cases of $2$ and $3$ sites in Tables \ref{tab:2sitesDiagrams} and \ref{tab:3sitesDiagrams}.
\begin{table}
\begin{center}
\begin{tabular}{l|l||c}\hline
$C$=Number of connections & Loop Diagrams & $S^z$ charge
\\ \hline
0 & $||$ & 1 \\
1 & $)|$ & -1 \\
2 & $()$ ; $))$ & 0
\end{tabular}
\end{center}
\caption{\label{tab:2sitesDiagrams} Diagrams for $L=2$ sites}
\end{table}
\begin{table}
\begin{center}
\begin{tabular}{l|l||c}\hline
$C$=Number of connections & Loop Diagrams & $S^z$ charge
\\ \hline
0 & $|||$ & $\f{3}{2}$\\
1 & $)||$ & $-\f{3}{2}$\\
2 & $))|$ ; $()|$ ; $|()$ & $\f{1}{2}$\\
3 & $)))$ ; $())$ ; $)()$ & $-\f{1}{2}$
\end{tabular}
\end{center}
\caption{\label{tab:3sitesDiagrams} Diagrams for $L=3$ sites}
\end{table}
It is easily seen that the action of the generators on any picture can never
decrease the number of connections and this fact allows us to block
diagonalize the Hamiltonian. Let us illustrate with the simple example at
$L=2$ sites. 

Acting on the basis:
\bea
\left( \begin{array}{c}
|| \\
)| \\
() \\
))
\end{array}
\right)
\eea
we have:
\bea
e_0&=&\left(
\begin{array}{llll}
0 & 0 & 0 & 0 \\
1 & \f{\sin \omega}{\sin(\omega+\gamma)} & 0 & 0 \\
0 & 0 & 0 & 0 \\
0 & 0 & 1 & \f{\sin \omega}{\sin(\omega+\gamma)}
\end{array}
\right) \\
e_1&=&\left(
\begin{array}{llll}
0 & 0 & 0 & 0 \\
0 & 0 & 0 & 0 \\
1 & 1 & 2 \cos \gamma & 1 \\
0 & 0 & 0 & 0
\end{array}
\right)
\eea
In \ref{sec:Faithfulness} we show that, in the case of $2$ generators,
for generic values of $\gamma$ and $\omega$ this
representation is faithful. However at some subset of the `critical' points
(\ref {eqn:boundaryTLcritical})  there exist additional relations between the
words of the 1BTL and it is no longer faithful. We believe this is also true for the general case.

In the link pattern representation the eigenstates of the diagrammatic charge are of course the individual
diagrams. The charge in this basis is therefore given by:
\bea
C&=&\left(
\begin{array}{llll}
0 & 0 & 0 & 0 \\
0 & 1 & 0 & 0 \\
0 & 0 & 2 & 0 \\
0 & 0 & 0 & 2
\end{array}
\right)
\eea
The Hamiltonian (\ref{eqn:TLlb}) is given by:
\bea
H^{lp}&=&-a e_0 -e_1=\left(
\begin{array}{cccc} 
\multicolumn{1}{c|}{0}& 0 & 0 & 0  \\ \cline{1-2}
\multicolumn{1}{c|}{-a} & \multicolumn{1}{c|}{-a \f{\sin \omega}{\sin(\omega+\gamma)}} & 0 & 0 \\\cline{2-4}
-1 & -1 & \multicolumn{1}{|c}{-2 \cos \gamma} & -1 \\
0 & 0 & \multicolumn{1}{|c}{-a} & -a \f{\sin \omega}{\sin(\omega+\gamma)}
\end{array}
\right)
\eea
where $a$ is given by (\ref{eqn:Definitionofa}). The fact that the generators can never break links gives the Hamiltonian
its lower block triangular structure. However it is important to note that the diagrammatic charge $C$ does \emph{not} commute with the 1BTL generators or with the Hamiltonian.

We can consider a `fake' Hamiltonian $\tilde{H}^{lp}$ which conserves
diagrammatic charge (this Hamiltonian was used in \cite{deGier:2003iu}):
\bea \label{eqn:FakeH}
\tilde{H}^{lp}&=& \left(
\begin{array}{cccc}
0  & 0 & 0 & 0 \\
0 & -a \f{\sin \omega}{\sin(\omega+\gamma)} & 0 & 0 \\
0 & 0 & -2 \cos \gamma & -1 \\
0 & 0 & -a & -a \f{\sin \omega}{\sin(\omega+\gamma)} 
\end{array}
\right)
\eea
This of course has the same eigenvalues as $H^{lp}$ but in general different eigenvectors. 

The dimension of the eigenspaces of $C$ (\ref{eqn:Diagchargespaces}) are
the same as that of the $S^z$ charge in the spin system if relate the
eigenvalues $m$ of $S^z$ where $-\f{L}{2}\le m\le \f{L}{2}$ to the values of
the diagrammatic charge in the following way:
\begin{itemize}
\item{$C$ even}
\bea
m=\f{L}{2}-C
\eea
\item{$C$ odd}
\bea
m=-\f{L}{2}+C-1
\eea
\end{itemize}
In fact in \ref{sec:Betheansatz} we prove using the Bethe ansatz
that the eigenvalues of $\tilde{H}^{lp}$ on the
sector with $S^z$ eigenvalue $m$ coincide with the
eigenvalues of $H^d$ in the sector with the same value of $m$. This is
illustrated in the final column of Tables
\ref{tab:2sitesDiagrams} and \ref{tab:3sitesDiagrams} for the cases of $L=2,3$.

As we shall show in the next section the appearance of indecomposable
structures related to `critical' (non-semisimple) algebras can be understood
in the case of
$H^{nd}$. However in the link pattern representation the Jordan structures are
unknown (see however \ref{sec:Similarities} and \ref{sec:Faithfulness} for the two site case). We have checked for different system sizes, by exact diagonalization, and found that for
$\gamma=\omega=\f{\pi}{3}$ there are no Jordan cell structures in $H^{lp}$. This implies
that for this case $H^{lp}$ can be brought by a similarity transformation to
$H^d$. This special case is of importance since the Hamiltonian:
\bea
H^{stochastic}=a+L-1+H^{lp}\left( \gamma=\f{\pi}{3};\omega=\f{\pi}{3}\right)
\eea
describes the time evolution of a fluctuating interface. The ground-state energy of $H^{stochastic}$ is zero for any number of sites and any value of $a \ge 0$ and has interesting combinatorial properties \cite{RefinedJanandVladimir}.
\section{Properties of the centralizer of the 1BTL algebra in the `non-diagonal' spin representation}
\setcounter{equation}{0}  
\subsection{Definition of the centralizer}
We have described up to now three Hamiltonians corresponding to
different representations of the 1BTL algebra (in the case of $H^d$
this is more of a conjecture than a statement - see \ref{sec:DiagonalRepn}). We
have also repeatedly mentioned the possibility of the appearance of
Jordan cells structures connected with the `critical' algebras (\ref{eqn:boundaryTLcritical}). We have also to keep in mind the occurrence of
degeneracies in the diagonal chain $H^d$ \cite{Alcaraz:1988vi,Grimm:1990dg,Grimm:1990gg}.

In this section we
are going to show that a centralizer of the 1BTL algebra in the
representation in which the generators are given by (\ref{eqn:TLgenerators})
and (\ref{eqn:e0}) is going to bring a new insight in these problems. This
centralizer was discovered by Doikou \cite{Doikou:2004km} but its properties and its
relevance have not yet been studied, we are going to do it in this
section.

For convenience, instead of the ferro-magnetic convention for the
generators $e_i$ used in (\ref{eqn:TLgenerators}), we will use the
anti-ferromagnetic notation:
\bea
e_i= -\half \left\{ \sigma^x_i \sigma^x_{i+1} + \sigma^y_i \sigma^y_{i+1} + \cos \gamma \sigma^z_i \sigma^z_{i+1}  - \cos \gamma + i \sin \gamma \left(\sigma^z_i - \sigma^z_{i+1} \right) \right\}
\eea
The expression of $e_0$ is also modified:
\bea
e_0&=&-\half \f{1}{\sin(\omega+\gamma)}\left( -i \cos \omega \sigma_1^z - \sigma_1^x - \sin \omega \right) \nonumber\\
&=&\half \f{1}{\sin(\omega+\gamma)}\left(
\begin{array}{cc}
i e^{-i \omega} & 1  \\
1 & -i e^{i \omega}
\end{array}
\right) \otimes {\bf 1} \otimes \cdots \otimes {\bf 1}
\eea
We have called this representation the `non-diagonal' representation.

The quantum group $U_q(SU(2))$, is generated by $S^{\pm},q^{\pm S^z}$ with the relations:
\bea \label{eqn:quantumgroup}
q^{S^z} S^{\pm} q^{- S^z}&=&q^{\pm} S^{\pm} \\
\left[ S^+, S^- \right] &=& \f{q^{2 S^z} - q^{-2 S^z}}{q-q^{-1}} \nonumber
\eea
and co-products given by:
\bea \label{eqn:qgcoproducts}
\Delta(S^{\pm})&=& q^{S^z} \otimes S^{\pm} + S^{\pm} \otimes q^{-S^z} \\
\Delta(q^{\pm S^z})&=& q^{\pm S^z} \otimes q^{\pm S^z} \nonumber
\eea
Using these we find that the action on the $SU(2)$ quantum spin chain is given by:
\bea 
q^{\pm S^z}&=&q^{\pm \half \sigma^3} \otimes \cdots \otimes q^{\pm \half \sigma^3} \\
S^{\pm}&=& \sum_i q^{\half \sigma^3} \otimes \cdots \otimes q^{\half \sigma^3}
\otimes \sigma_i^{\pm} \otimes q^{-\half \sigma^3} \otimes \cdots \otimes
q^{-\half \sigma^3}  \nonumber
\eea
As is well known \cite{Pasquier:1989kd}, the generators of $U_q(SU(2))$, (with $q=e^{i
  \gamma}$) commute with the $e_i$'s.

The generators $S^z$, $S^+$ and $S^-$ do not commute however with
$e_0$. Doikou \cite{Doikou:2004km} has shown that in the `non-diagonal'  representation, the 1BTL
algebra has a centralizer $X$:
\bea
\left[ X,e_i \right]=0 \quad \quad \left[ X,e_0 \right]=0
\eea
which has the following expression:
\bea \label{eqn:Centralizer}
X=\f{1}{2 \sin (\gamma + \omega)} \left\{e^{- \half i \gamma} S^+ q^{-S^z} + e^{+ \half i \gamma} S^- q^{-S^z} - \f{\cos \omega}{\sin \gamma} \left( q^{-2S^z}-1 \right) \right\}
\eea
where we have used the definition (\ref{eqn:quantumgroup}) and
 (\ref{eqn:qgcoproducts}) of the quantum group generators. The normalization
 factor in (\ref{eqn:Centralizer}) is, as it going to be seen, a very convenient one.
 One can see the expression of $X$ as a `co-product' form (very different
from (\ref{eqn:qgcoproducts})):
\bea \label{eqn:coproductX}
\Delta(X)=1 \otimes X + X \otimes q^{-2S^z}
\eea
It is clearly obvious that $X$ commutes with the bulk \TL generators as it is constructed from $U_q(SU(2))$ quantum group generators. Using the co-product and the action on one site:
\bea \label{eqn:Centralizer1site}
X_1&=&\f{1}{2 \sin(\gamma+\omega)}\left(
\begin{array}{cc}
\f{i e^{-i \gamma/2} \cos \omega}{\cos \f{\gamma}{2}} & 1 \\
1 & \f{-i e^{i \gamma/2} \cos \omega}{\cos \f{\gamma}{2}}
\end{array}
\right)\nonumber\\
&=& e_0+ \f{\sin(\half \gamma-\omega)}{2 \cos (\half \gamma)
  \sin(\gamma+\omega)} {\bf 1}
\eea
one can easily see that it also commutes with the boundary generator $e_0$.

We shall show in the next section that the centralizer $X$ becomes
indecomposable precisely in the `critical' cases (\ref{eqn:boundaryTLcritical}). This implies (see \ref{sec:Jordan}) that degeneracies are induced in $H^{nd}$ and
implicitly in $H^d$ and $H^{lp}$. The results of \ref{sec:Similarities} for
$2$ sites and various numerical tests show that in the `critical' cases
$H^{nd}$ always has the same indecomposable structure as $X$. In contrast the
Hamiltonians $H^{lp}$ and $H^d$, 
although having the same spectrum and degeneracies as
$H^{nd}$, did not always have Jordan structure at the critical points. We
found that the Hamiltonian $H^{lp}$ had Jordan structure at some subset of the
`critical' points whereas $H^d$ was always fully diagonalizable.
\subsection{The spectrum and eigenfunctions of $X$: The ${\bf Q}$-basis}
\label{sec:Qbasis}
We shall now calculate the eigenvalues and eigenvectors of the centralizer
(\ref{eqn:Centralizer}).

Firstly from diagonalizing the action of $X$ on a one site system (\ref{eqn:Centralizer1site}) we find its two eigenvalues can be written as:
\bea \label{eqn:Xeigenvalues1site}
\f{\sin (-\half \gamma) \sin(-\half \gamma + \omega)}{\sin \gamma \sin (\gamma+\omega)}, \f{\sin (\half \gamma) \sin(\half \gamma + \omega)}{\sin \gamma \sin (\gamma+\omega)}
\eea
We shall denote the operator $X$ acting on an $L$ site chain by
$X^{(L)}$. Motivated by exact diagonalization at a low number of sites we
found that the eigenvalues of $X^{(L)}$ took the following form:
\bea  \label{eqn:Xeigenvaluesconj}
\f{\sin Q \gamma \sin(Q \gamma + \omega)}{\sin \gamma \sin (\gamma+\omega)} \quad {\rm~ with~ degeneracy:} \quad \left(\begin{array}{c} L\\ \f{L}{2}-Q  \end{array} \right)
\eea
with $Q=-\f{L}{2},\cdots,\f{L}{2}$. The degeneracy given in
(\ref{eqn:Xeigenvaluesconj}) is for generic
values of the parameters $\gamma$ and $\omega$. As we shall explain later this
degeneracy becomes enhanced in the `critical' cases (\ref{eqn:boundaryTLcritical}). The one site results (\ref{eqn:Xeigenvalues1site}) are for $Q=-\half,\half$. The fact that the degeneracy is exactly that of an $S^z$ operator suggests that we index the $2^L$ eigenvectors by the vector label ${\bf Q}=(Q_1;Q_2;\cdots;Q_L)$ where $Q_i=\pm \half$. The degeneracy is immediately explained if we take the eigenvalues to be:
\bea \label{eqn:Xeigenvalues}
\lambda^{(L)}({\bf Q})=\f{\sin Q \gamma \sin(Q \gamma + \omega)}{\sin \gamma \sin (\gamma+\omega)} \quad  \quad {\rm with:} \quad Q=\sum_{i=1}^{L} Q_i
\eea
We shall now use the co-product of $X$ (\ref{eqn:coproductX}) to prove this by induction and to find a general formula for the eigenvectors.

The eigenvectors ${\bf v}^{(L)}({\bf Q}^{(L)})$ are vectors in the spin basis satisfying:
\bea
X^{(L)} {\bf v}^{(L)}({\bf Q})=\lambda^{(L)}({\bf Q}){\bf v}^{(L)}({\bf Q})
\eea
We now form the most general eigenvectors of $X^{(L+1)}$. These are indexed by a $L+1$ dimensional vector ${\bf R}=(R_1;R_2;\cdots;R_{L+1})$:
\bea \label{eqn:geneigenvector}
{\bf v}^{(L+1)}({\bf R})&=&\sum_{{\bf Q}} \left\{ a({\bf R},{\bf Q}) {\bf v}^{(L)}({\bf Q}) \otimes \uparrow + b({\bf R},{\bf Q}) {\bf v}^{(L)}({\bf Q}) \otimes \downarrow \right\}
\eea
Then using the co-product (\ref{eqn:coproductX}) we have:
\bea
X^{(L+1)} {\bf v}^{(L+1)}({\bf R})&=&\sum_{{\bf Q}} \left\{ {\bf v}^{(L)}({\bf Q}) \otimes \uparrow \left[ a({\bf R},{\bf Q})A +b({\bf R},{\bf Q})B+q^{-1}a({\bf R},{\bf Q}) \lambda^{(L)}({\bf Q})\right] \right. \nonumber \\
&& \left. + {\bf v}^{(L)}({\bf Q}^{(L+1)}) \otimes \downarrow \left[ a({\bf R},{\bf Q})C+ b({\bf R},{\bf Q})D+q b({\bf R},{\bf Q})\lambda^{(L)}({\bf Q}) \right] \right\}\nonumber\\
&=&  \lambda^{(L+1)}({\bf R}) \sum_{{\bf Q}} \left\{ a({\bf R},{\bf Q}) {\bf v}^{(L)}({\bf Q}) \otimes \uparrow + b({\bf R},{\bf Q}) {\bf v}^{(L)}({\bf Q}) \otimes \downarrow \right\}
\eea
where we abbreviate the action of $X$ on a single site (\ref{eqn:Centralizer1site}) by:
\bea \label{eqn:abbreviationofX}
X \left( 
\begin{array}{c} 
\uparrow \\
\downarrow
\end{array} \right)
= \left(\begin{array}{cc} 
A & B \\
C & D 
\end{array} \right)
\left(
\begin{array}{c} 
\uparrow \\
\downarrow
\end{array} \right)
\eea
Comparing terms we find:
\bea \label{eqn:recurrence}
\lambda^{(L+1)}({\bf R}) a({\bf R},{\bf Q})&=&a({\bf R},{\bf Q})A+b({\bf R},{\bf Q})B+q^{-1} a({\bf R},{\bf Q}) \lambda^{(L)}({\bf Q}) \\
\lambda^{(L+1)}({\bf R}) b({\bf R},{\bf Q})&=&a({\bf R},{\bf Q})C+b({\bf R},{\bf Q})D+q b({\bf R},{\bf Q}) \lambda^{(L)}({\bf Q}) \nonumber
\eea
Now we can eliminate $a({\bf R},{\bf Q}),b({\bf R},{\bf Q})$:
\bea \label{eqn:quadraticforevals}
\left( \lambda^{(L+1)}({\bf R})-A - q^{-1} \lambda^{(L)}({\bf Q})\right) \left( \lambda^{(L+1)}({\bf R})-D - q \lambda^{(L)}({\bf Q})\right)=B C 
\eea
Inserting the solution $\lambda^{(L)}({\bf Q})$ valid for $L$ sites (\ref{eqn:Xeigenvalues}) we find two solutions to (\ref{eqn:quadraticforevals}) for $\lambda^{(L+1)}({\bf R})$:
\bea \label{eqn:newsolns}
\lambda^{(L+1)}({\bf R})=\left\{ \begin{array}{cc}
\f{\sin ((Q+\half) \gamma) \sin((Q+\half) \gamma + \omega)}{\sin \gamma \sin (\gamma+\omega)}\\
\f{\sin ((Q-\half) \gamma) \sin((Q-\half) \gamma + \omega)}{\sin \gamma \sin (\gamma+\omega)}
\end{array} \right.
\eea
Therefore we see that:
\bea
\sum_{i=1}^{L+1} R_i=\sum_{i=1}^{L} Q_{i} \pm \half
\eea
Now it is important to note that this constraint still allows considerable flexibility in the definition of the vector ${\bf R}=(R_1;\cdots;R_{L+1})$. In particular it can be satisfied by choosing ${\bf R}=\left({\bf Q}; R_{L+1}\right)$ with the notation that the first $L$ components of ${\bf R}$ are given by ${\bf Q}$ and the final one by $R_{L+1}=\pm \half$. The motivation is that the similarity of ${\bf Q}$ to ${\bf S}^z$ might extend to allow the eigenvectors to be written as tensor products. At this point however this should be regarded as an ansatz whose ultimate justification will come when we are able to use it to write down a complete set of linearly independent eigenvectors. With this choice the sum (\ref{eqn:geneigenvector}) reduces to a single term:
\bea
{\bf v}^{(L+1)}({\bf Q};R_{L+1})&=&a({\bf Q};R_{L+1}) {\bf v}^{(L)}({\bf Q}) \otimes \uparrow + b({\bf Q};R_{L+1}) {\bf v}^{(L)}({\bf Q}) \otimes \downarrow \nonumber \\
&=& {\bf v}^{(L)}({\bf Q}) \otimes \left[ a({\bf Q};R_{L+1}) \uparrow + b({\bf Q};R_{L+1}) \downarrow \right]
\eea
where we have, by an abuse of notation, written $a({\bf Q};R_{L+1})$ for the previous $a({\bf R},{\bf Q})$ with ${\bf R}=\left({\bf Q}; R_{L+1}\right)$ and similarly for $b({\bf Q};R_{L+1})$.

In other words the eigenstates can be written in a \emph{tensor product}
form. This is an important simplification and allows us to write them in a
general form. To see this let us return to (\ref{eqn:recurrence}) and, knowing
$\lambda^{(L)}({\bf Q})$ and $\lambda^{(L+1)}({\bf R})$ from
(\ref{eqn:Xeigenvalues}) and (\ref{eqn:newsolns}), solve for $a({\bf Q};R_{L+1})$ and $b({\bf Q};R_{L+1})$. Firstly, up to normalization of the eigenfunction, only the ratio of these two functions is significant:
\bea
\f{a({\bf Q};R_{L+1})}{b({\bf Q};R_{L+1})}= \f{\lambda^{(L+1)}({\bf Q};R_{L+1}) -D -q \lambda^{(L)}({\bf Q})}{C}
\eea
Inserting the formulae for $C,D$ defined in (\ref{eqn:abbreviationofX}) and the eigenvalues (\ref{eqn:Xeigenvalues}) one gets the following very simple expression:
\bea \label{eqn:phasefactor}
\f{a({\bf Q};R_{L+1})}{b({\bf Q};R_{L+1})}=\left\{
\begin{array}{c}
i e^{-2i \gamma Q - i \omega} \quad R_{L+1}=+\half \\
i e^{2i \gamma Q + i \omega} \quad R_{L+1}=-\half
\end{array} \right.
\eea
These can be combined into a single expression:
\bea
\f{a({\bf Q};R_{L+1})}{b({\bf Q};R_{L+1})}= i e^{-4i \gamma Q R_{L+1} - 2i \omega R_{L+1}}
\eea
As the R.H.S. is finite we can choose the normalization $b({\bf
  Q},R_{L+1})=1$. Then we obtain:
\bea
{\bf v}^{(L+1)}({\bf Q},R_{L+1})=v^{(L)}({\bf Q}) \otimes \left[ i e^{ -4i \gamma Q R_{L+1} - 2i \omega R_{L+1}} \uparrow + \downarrow \right]
\eea
where $Q=Q_1+\cdots+Q_L$. This equation defines a recurrence relation for the eigenvectors and can easily be solved. 

The final result for the eigenvectors is:
\bea \label{eqn:eigenvectors}
{\bf v}^{(L)}({\bf Q})&=&\left[ i e^{ - 2i \omega Q_{1}} \uparrow + \downarrow \right] \otimes \left[ i e^{ -4i \gamma Q_1 Q_{2} - 2i \omega Q_{2}} \uparrow + \downarrow \right] \\
&& \otimes \left[ i e^{ -4i \gamma (Q_1+Q_2) Q_{3} - 2i \omega Q_{3}} \uparrow + \downarrow \right] \otimes \left[ i e^{ -4i \gamma (Q_1+Q_2+Q_3) Q_{4} - 2i \omega Q_{4}} \uparrow + \downarrow \right] \cdots \nonumber\ \\
&& \otimes \left[ i e^{ -4i \gamma (Q_1+Q_2+\cdots+Q_{L-1}) Q_{L} - 2i \omega Q_{L}} \uparrow + \downarrow \right] \nonumber
\eea
where ${\bf Q}=(Q_1,Q_2,\cdots,Q_L)$. We shall refer to this as the ${\bf Q}$-basis. The completeness of this basis will be discussed in the next subsection.

We have verified at one and two sites, by explicit diagonalization, that the
eigenvectors are indeed given by the above form. Explicitly they are:
\begin{itemize}
\item{One site}
\bea \label{eqn:Onesiteresults}
\begin{array}{ll}
Q=+\half: & ie^{ - i \omega} \uparrow + \downarrow \\
Q=-\half: & ie^{  i \omega} \uparrow + \downarrow
\end{array}
\eea
These states are linearly independent except when $w=0$.
\item{Two sites}
\bea \label{eqn:Twositeresults}
\begin{array}{ll}
{\bf Q}=(+\half;+\half): 
& -e^{-i \gamma - 2i \omega} \uparrow \uparrow + i e^{ - i \omega} \uparrow \downarrow + i
e^{ -i \gamma - i \omega } \downarrow \uparrow + \downarrow \downarrow \\
{\bf Q}=(+\half;-\half):& 
-e^{i \gamma} \uparrow \uparrow + i e^{ - i \omega} \uparrow \downarrow + i
e^{ i \gamma + i \omega } \downarrow \uparrow + \downarrow \downarrow \\
{\bf Q}=(-\half;+\half):& 
-e^{i \gamma} \uparrow \uparrow + i e^{  i \omega} \uparrow \downarrow + i
e^{ i \gamma - i \omega } \downarrow \uparrow + \downarrow \downarrow \\
{\bf Q}=(-\half;-\half):& 
-e^{-i \gamma+2 i \omega} \uparrow \uparrow + i e^{  i \omega} \uparrow \downarrow + i
e^{ -i \gamma + i \omega } \downarrow \uparrow + \downarrow \downarrow
\end{array}
\eea
These states are linearly independent except when $\omega=-\gamma,0, \gamma$. 
\end{itemize}
The spectrum of $X$ (\ref{eqn:Xeigenvaluesconj}) has additional degeneracies
in the cases when $\lambda^{(L)}(Q)=\lambda^{(L)}(Q')$ has non-trivial
solutions. This implies that we have solutions to one, or both, of the equations:
\bea \label{eqn:1BTLdegeneracies}
\quad (Q+Q')\gamma +\omega &=& \pi {\bf Z}
\eea
\bea \label{eqn:QGdegeneracies}
(Q-Q')\gamma &=& \pi {\bf Z}
\eea
As we shall see in the next subsection the first equation
(\ref{eqn:1BTLdegeneracies}) is related to the `critical' points of the 1BTL
algebra. The role of the second one (\ref{eqn:QGdegeneracies}) which is
typical when a $U_q(SU(2))$ symmetry plays a role, is not obvious in
the present context. Taking $\omega$ generic and $q = e^{i \gamma}$ a root of
unity, one obtains degeneracies in the spectrum of $X$. However there are no
Jordan cell structures in the expression of $X$. We would like to
stress that the completeness of the ${\bf Q}$-basis (\ref{eqn:eigenvectors})
is not affected by (\ref{eqn:QGdegeneracies}). Simple examples suggest for
$q$ a root of unity there exist additional centralizers which do not have
superfluous degeneracies. We hope to come back to this problem in another publication.
\subsection{Truncated Bratelli diagrams and the critical points}
\label{sec:Bratelli}
\begin{figure}
\centering
\includegraphics[width=8 cm]{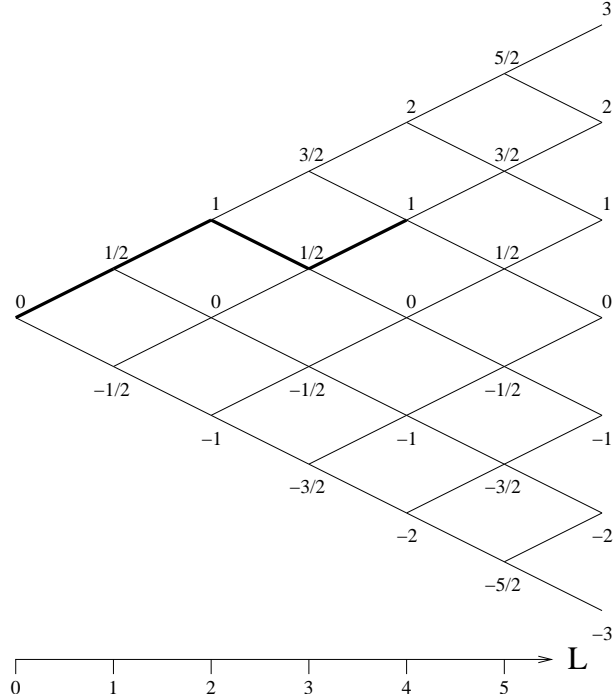}
\caption{\label{fig:fullbratelli} Full Bratelli Diagram. The system size, $L$,
  is given on the horizontal axis. The path
  corresponding to the eigenstate ${\bf v}^{(4)}(\half,\half,-\half,\half)$ is
  shown in bold.}
\end{figure}

The space of eigenvectors, with their values of ${\bf Q}$ can be encoded in a
Bratelli diagram (see Figure \ref{fig:fullbratelli}).
From each different path on the diagram one reads off the values of $Q_i$ and
gets an eigenvector from (\ref{eqn:eigenvectors}). As there are two choices at each point ($Q_i=\pm \half$) it
is obvious that the space of solutions (\ref{eqn:eigenvectors})
has dimension $2^L$. For a system of size $L$ the degeneracy corresponding to a given value
of $Q$ is given by the binomial coefficient in
(\ref{eqn:Xeigenvaluesconj}). In terms of the Bratelli diagram it is the
number of paths that start from the far left side at $0$ and reach that point.

The set of eigenvectors for a system of size $L$ may not always be linearly
independent. The only problem in the construction of the basis is when the two possible expressions in (\ref{eqn:phasefactor}) coincide. In this case we have:
\bea
e^{-2i \gamma Q - i \omega}= e^{+2i \gamma Q + i \omega} 
\eea
and therefore we have:
\bea \label{eqn:exceptionalQ}
2 \gamma Q +\omega=\pi {\bf Z}
\eea
This is the generalization of the results for $1$ and $2$ sites:
(\ref{eqn:Onesiteresults}) and (\ref{eqn:Twositeresults}). As $Q$ takes integer and
half-integer values (\ref{eqn:exceptionalQ}) gives exactly the `critical' cases of the one-boundary \TL algebra (\ref{eqn:boundaryTLcritical}).

In these critical cases the basis breaks down. If from a particular state we
can only produce one new eigenvector in (\ref{eqn:phasefactor}) rather than
two then it means
that a complete basis of eigenvectors cannot be constructed - in other words
the centralizer is becoming of Jordan form. In this case
one can form a reduced space in which we simply remove all such
states.

Let us illustrate with the example $\gamma=\f{\pi}{5}, \omega=2
\pi/5$. The action of
the centralizer on the states $V_Q$ of different $Q$ number is given by:

\begin{center}
\begin{tabular}{c|ccccccccc}
 & \multicolumn{9}{c}{Value of $Q$} \\
System size $(L)$ & $-2$ & $-\f{3}{2}$ & $-1$ &  $-\half$& $0$ & $\half$
& $1$ & $\f{3}{2}$ & $2$\\
\hline
1 & - & - & -& 1 & - & 1 & -& - & -\\
2 & - & - & 1& - & 2 & - & 1& - & -\\
3 & - & 1$^{*}$ & -& 3$^{*}$ & - & 3 & -& 1 & - \\
4 & 1$^{*}$ & - & 4 & - & 6$^{*}$ & - & 4$^{**}$ & - & 1$^{**}$ \\
\hline
Eigenvalues of $X$ & 0 & $\f{-1}{\sqrt{5}}$ & $\f{1-\sqrt{5}}{2}$ &$\f{-1}{\sqrt{5}}$
& 0 &$1-\f{1}{\sqrt{5}}$ & 1 & $\f{5+3 \sqrt{5}}{10}$ & 1
\end{tabular}
\end{center}

In the cases in which the eigenvalues of the centralizer take different values
in each $Q$ sector there is a complete set of $2^L$ eigenvectors given by
(\ref{eqn:eigenvectors}). However, as one can see from the above table, some
of the different $Q$ sectors give the same eigenvalues of the centralizer $X$
and, by explicit diagonalization, one finds that different
sectors begin to mix (indicated by $^{*}$ and $^{**}$). At $L=3$ we find
that the 
$4$ states from $Q=-\f{3}{2}$ and $Q=-\half$ combine into two eigenvectors and a single two dimensional Jordan cell. At $L=4$ the $Q=-2$ and $Q=0$ states form $5$ eigenvectors and a single two dimensional Jordan cell; the $Q=1$ and $Q=2$ states form $3$
eigenvectors and 
a single indecomposable representation. 

The fact that the centralizer cannot be completely diagonalized leads to
degeneracy in the Hamiltonian $H^{nd}$ (see
\ref{sec:Jordan}). However even for larger system sizes (we checked up
to $L=6$) we found that the indecomposable structures always broke into pairs
and therefore the degeneracy in the Hamiltonian was never more than doublets. Using the
spectral equivalence this gives us an understanding of the appearance of
singlets and doublets in the diagonal chain $H^d$. This is interesting because
in the numerical work on the diagonal chain in the case $\omega=-\gamma$ 
\cite{Alcaraz:1988vi,Grimm:1990dg,Grimm:1990gg} the minimal models were
obtained by disregarding the doublets. We now have an understanding of this
numerical prescription - in the `non-diagonal'
representation it simply corresponds to discarding the indecomposable sector of the 1BTL! 

From the condition (\ref{eqn:exceptionalQ}) we
see that the first occurrence of indecomposable structure is for $Q=-1,\f{3}{2}$. This is seen in the
above table as these states begin to mix. We can try to truncate the space of
states by removing 
\emph{all} the states created from $Q=-1$ and $\f{3}{2}$. This leaves us with
the diagram:
\begin{center}
\includegraphics[width=8 cm]{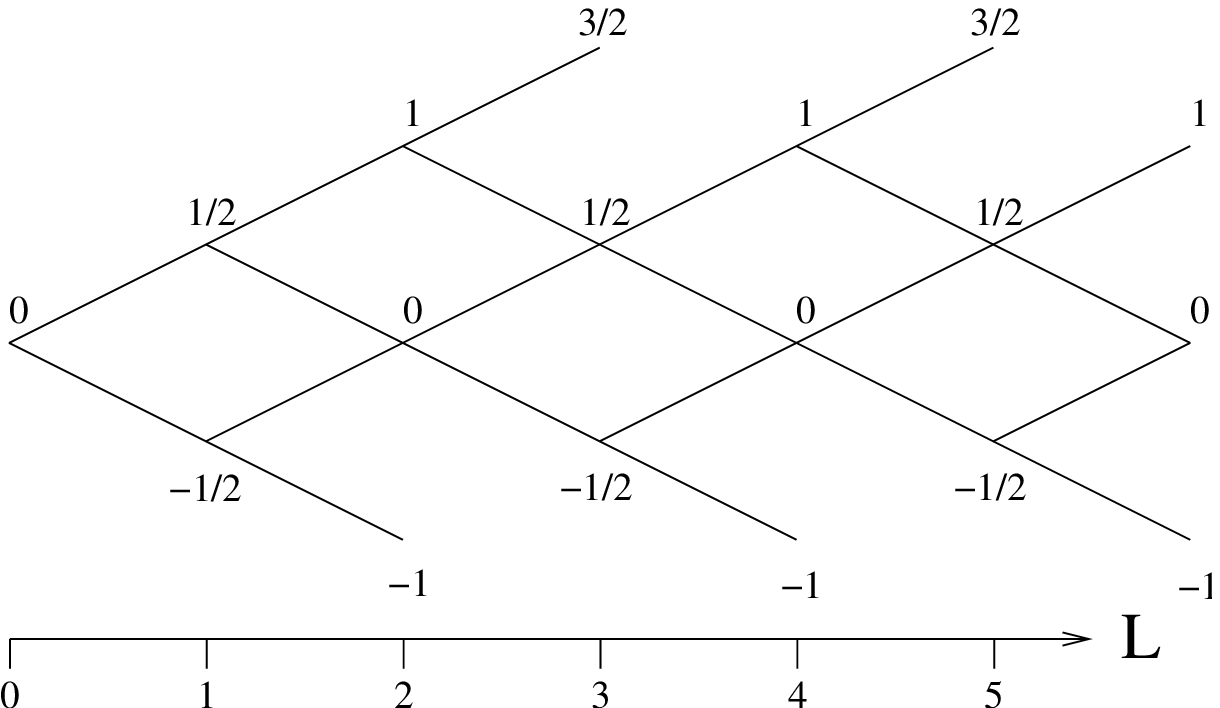}
\end{center}
On this
truncated space the centralizer is completely diagonalizable. These states
form the irreducible representations of the 1BTL algebra in the XXZ
representation. The degeneracy of a particular state can be read off as before
by counting the number of paths leading to that point. In the case of
$\gamma=\f{\pi}{5}, \omega=\f{3 \pi}{5}$ the truncated Bratelli diagram is given by:
\begin{center}
\includegraphics[width=8 cm]{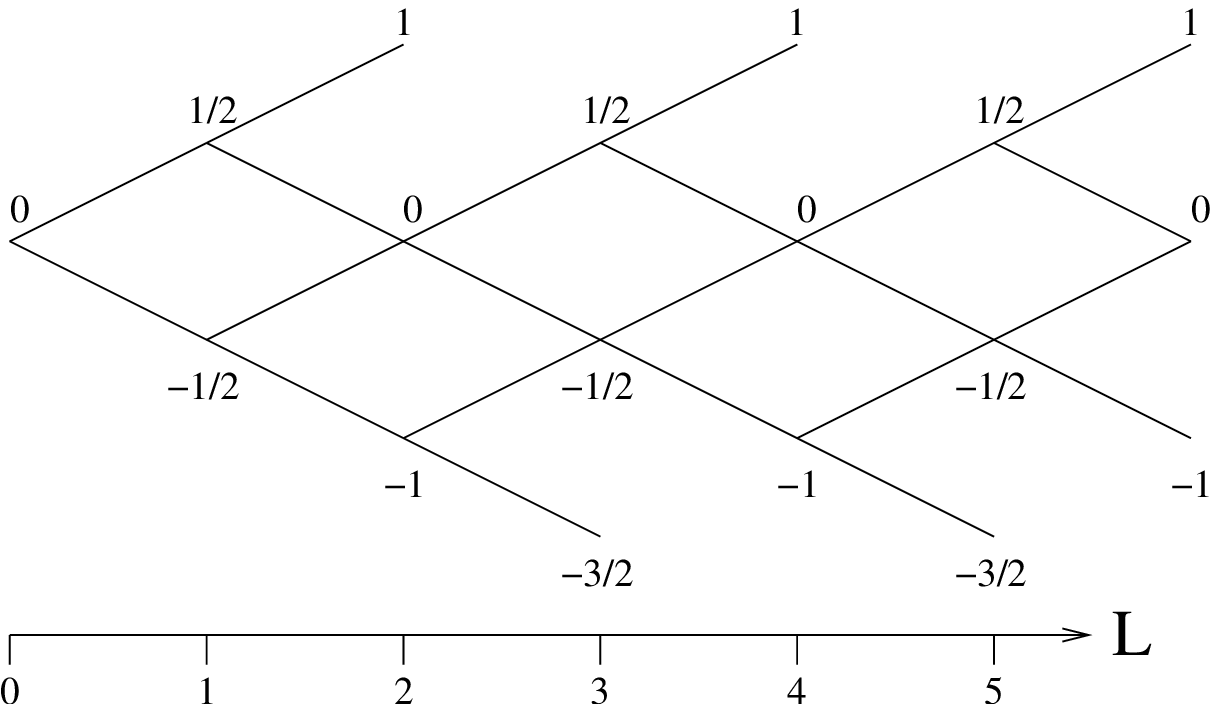}
\end{center}
The
truncated Bratelli diagrams presented here were derived from the properties of the centralizer
X. For a different approach, see the papers of Martin et
al. \cite{Martin:1992td,Martin:1993jk,MartinWoodcockI,MartinWoodcockII}.
\subsection{Action of 1BTL generators in ${\bf Q}$-basis}
For generic values of the parameters the basis of eigenvectors
(\ref{eqn:eigenvectors}) can be constructed and used to diagonalize the
Hamiltonian. The ${\bf Q}$-basis is written in terms of states:
\bea
\left| Q_1 ; Q_2 \cdots ;  Q_L \right>
\eea
The expressions for these in terms of the spin basis are given in
(\ref{eqn:eigenvectors}). As the action for $e_0$ and $e_i$ are known in the
spin basis we can work out their action on the ${\bf Q}$-basis. For $e_0$ we have:
%
%
%
\bea \label{eqn:e0inQbasis}
e_0 \left| \half ; Q_2 \cdots ;  Q_L \right> &=& \f{\sin
  w}{\sin(\omega+\gamma)} \left| \half ; Q_2 \cdots ;  Q_L \right> \\
e_0 \left| -\half ; Q_2 \cdots ;  Q_L \right> &=& 0 \nonumber
\eea
For $e_i$ we have:
%
%
%
\bea \label{eqn:eiinQbasis}
e_i \left|  \cdots ; Q_{i-1}; \half ; \half ; Q_{i+2} ; \cdots \right> &=&0
\nonumber \\
e_i \left|  \cdots ; Q_{i-1}; \half ; -\half ; Q_{i+2} ; \cdots \right> &=&
\alpha \left|  \cdots ; Q_{i-1}; \half ; - \half ; Q_{i+2} ; \cdots ; \right> \nonumber
\\
&&-\alpha \left| \cdots ; Q_{i-1}; -\half ; \half ; Q_{i+2} ; \cdots \right>\\
e_i \left|  \cdots ; Q_{i-1}; -\half ;  \half ; Q_{i+2} ; \cdots \right> &=&
-  \beta \left|  \cdots ; Q_{i-1}; \half ; - \half ; Q_{i+2} ; \cdots ; \right>\nonumber
\\
&&  \beta \left| \cdots ; Q_{i-1}; -\half ; \half ; Q_{i+2} ; \cdots
\right>\nonumber \\
e_i \left|  \cdots ; Q_{i-1}; -\half ; -\half ; Q_{i+2} ; \cdots \right> &=&0 \nonumber 
\eea
where:
\bea \label{eqn:AlphaandBeta}
\alpha&=& \f{\sin(2 \gamma \tilde{Q} + \omega + \gamma)}{\sin(2 \gamma \tilde{Q} + \omega)} \nonumber\\
\beta&=& \f{\sin(2 \gamma \tilde{Q} + \omega - \gamma)}{\sin(2 \gamma \tilde{Q} + \omega)}
\eea
and both $\alpha$ and $\beta$ depend on the previous ${\bf Q}$ spins through $\tilde{Q}=Q_1+Q_2+\cdots+Q_{i-1}$ and therefore the $e_i$'s act non-locally.

Within the ${\bf Q}$-basis one can see from (\ref{eqn:e0inQbasis}) and
(\ref{eqn:eiinQbasis}) that both $e_0$ and the $e_i$'s act within sectors of
a given value of $Q$. This was to be expected as the ${\bf Q}$-basis is
a basis for the centralizer $X$.

In the ${\bf Q}$-basis the highest weight vector $\left|\half;\half;\cdots;\half
\right>$ and lowest weight vector $\left|-\half;-\half;\cdots;-\half \right>$ are eigenstates of all the 1BTL generators:
\bea
e_0 \left|\half;\half;\cdots;\half \right>&=& \f{\sin w}{\sin(\omega+\gamma)} \left|\half;\half;\cdots;\half \right>
\\
e_i \left|\half;\half;\cdots;\half \right>&=&0 \\
e_0 \left|-\half;-\half;\cdots;-\half \right>&=&0 \\
e_i \left|-\half;-\half;\cdots;-\half \right>&=&0
\eea
This implies that $H^{nd}$ has at least two eigenvalues that are independent of $L$. In some `critical' cases in which we have Jordan cell structures the highest and/or lowest weight state might not belong to the truncated ${\bf Q}$-basis.
\subsection{Two simple cases}
In this subsection we shall show that for two of the exceptional cases, namely $\omega=\pm \gamma$, the action of
the `master' Hamiltonian (\ref{eqn:TLlb}) has linear (or no) dependence on
the parameter $a$ in  the truncated sector. These phenomena were first
observed in numerical results \cite{Alcaraz:1988vi}. The case $\omega=-
\gamma$ was discussed more recently using completely different methods in \cite{Belavin:2003id}.
\begin{itemize}
\item{$\omega=-\gamma$}

For $L > 1$ the truncated sector is made purely from states:
\bea
\left| -\half;Q_2;\cdots;Q_L \right>
\eea
using the action of $e_0$ (\ref{eqn:e0inQbasis}) given in the previous section we find:
\bea
H \left| -\half;Q_2;\cdots;Q_L \right>=-\sum_{i=1}^{L-1}e_i \left| -\half;Q_2;\cdots;Q_L \right>
\eea
and therefore the energy levels in the truncated sector do not depend on
$\delta$~!
\item{$\omega= \gamma$}

For $L > 1$ the truncated sector is made purely from states:
\bea
\left| \half;Q_2;\cdots;Q_L \right>
\eea
using the action of $e_0$ given in the previous section
(\ref{eqn:e0inQbasis}) we find:
\bea
H \left| \half;Q_2;\cdots;Q_L \right>=\left(-a \f{\sin \gamma}{\sin(2 \gamma)} -\sum_{i=1}^{L-1}e_i\right) \left| \half;Q_2;\cdots;Q_L \right>
\eea
and therefore the energy levels in the truncated sector have only a linear
dependence on $a$. In this sector the
energy \emph{differences} between the excited states and ground state are independent of $\delta$. 
\end{itemize}
These results are interesting because in \cite{Alcaraz:1988vi} the spectrum of the
Potts models with free boundary conditions was numerically found in the diagonal chain with
$\omega=-\gamma$ by discarding the doublets and keeping the singlets. The
energy of these singlet states did not depend on the value of $\delta$. 

\section{Conclusions and open questions}
The surprising result of this paper is that two XXZ Hamiltonians $H^{nd}$
(\ref{eqn:Hnd}) and $H^d$ (\ref{eqn:Hd}), with the same bulk terms but very
different boundary conditions, and one Hamiltonian $H^{lp}$ defined in the
vector space of link patterns (see Section $3$), all have the same
spectrum. The proof is given in \ref{sec:Betheansatz} using the Bethe
ansatz. An interesting result of these calculations is that a
so-called `on the wrong side of the equator' problem shows up in
$H^d$, contrary to the periodic case \cite{Baxter:2001sx}.  

All three Hamiltonians are obtained considering a `master' Hamiltonian
(\ref{eqn:TLlb}) defined in terms of the generators of the one-boundary Temperley-Lieb
(1BTL) algebra (see (\ref{eqn:TL}) and (\ref{eqn:BoundaryTL})). This algebra
depends on two parameters $\gamma$ and $\omega$. The third parameter $\delta$
which appears in the coefficients of the boundary terms of $H^{nd}$ and $H^d$
does not appear in the definition of the algebra but is related to a
coefficient of the boundary generator in the `master' Hamiltonian.

The three Hamiltonians $H^{nd}$, $H^d$ and $H^{lp}$ correspond to different
representations of the 1BTL algebra. The representations used for $H^{nd}$ and
$H^{lp}$ are known ones \cite{MartinWoodcockII,Doikou:2002ry,JanReview}. The existence of a representation which should give
$H^d$, is less obvious. In the example of two sites, this representation is
given in \ref{sec:Similarities}. It is shown that the two generators commute
with the local charge $S^z$ (\ref{eqn:Sz}). We conjecture that such a
representation exists for any number of generators. This conjecture is not
so far-fetched since, as is going to be shown in \cite{WorkInProgress}, a representation of the 1BTL algebra in which
all the generators commute with $S^z$ exists. In this representation the bulk
generators have the standard form (\ref{eqn:TLgenerators}) but the boundary
generator is non-local. What remains to be shown is that a similarity
transformation brings us to the form (\ref{eqn:Hd}) of $H^d$. So far we have
only found this similarity transformation for small lattices.

The existence of `critical' points, at which  certain relations between the
parameters $\gamma$ and $\omega$ are satisfied (\ref{eqn:boundaryTLcritical}) was pointed out by
Martin et al. \cite{Martin:1992td,Martin:1993jk,MartinWoodcockI,MartinWoodcockII}. In these cases the 1BTL algebra possesses indecomposable
representations. We have to stress that the indecomposable structures in the
1BTL algebra have nothing to do with those observed in the bulk \TL algebra. 

The fact that the three Hamiltonians have the same spectra does not always
imply that they are equivalent. This is due to the fact that they can possess
different Jordan cell structures. In the simple example of two generators it is
shown in \ref{sec:Similarities} that there exist similarity transformations
relating the three Hamiltonians at generic values
of the parameters $\gamma$ and $\omega$. These transformations break down in
cases in which one Hamiltonian has Jordan form but the other does not. These
cases occur at (some subset of) the `critical' points of the 1BTL. The
question of whether a Hamiltonian in a given representation has Jordan cell form or is fully
diagonalizable is linked to the question of the faithfulness of the
representation (\ref{sec:Faithfulness}). 

In the case of $H^{nd}$, constructed using the `non-diagonal' representation,
a very good 
insight into the structure of indecomposable representations and
degeneracies can be obtained using the centralizer
$X$ (\ref{eqn:Centralizer}) discovered by Doikou \cite{Doikou:2004km}. In this representation the centralizer commutes
with the generators of the 1BTL. Although for generic values of $\gamma$ and $\omega$ it is
diagonalizable, in the
`critical' cases it possesses an indecomposable structure. We have computed
the spectrum of $X$ and found a basis in which 
$X$ is diagonal (for generic values
of $\gamma$ and $\omega$). This is the ${\bf Q}$-basis defined in section \ref{sec:Qbasis}. In the ${\bf Q}$-basis, one can define an
operator having the same eigenvalues (denoted by $Q$) and degeneracies as
those seen in
the spectrum of $S^z$. The generators of the 1BTL algebra conserve $Q$. If certain
relations between $\omega$ and $\gamma$ are satisfied (precisely the
`critical' cases found
by Martin et al. \cite{Martin:1992td,Martin:1993jk,MartinWoodcockI,MartinWoodcockII}), $X$ is not fully diagonalizable and the  ${\bf Q}$-basis is not
complete. It turns out the only Jordan cell structures that can appear are
\emph{two} dimensional. This is observation is based on many numerical examples.  We also show how to
obtain a reduced vector space (without Jordan cells) using truncated Bratelli
diagrams. 

What did we learn from the existence of the centralizer of the `non-diagonal'
representation? Firstly in the generic cases $H^{nd}$ can be block
diagonalized using the eigenspaces of $X$. Secondly in the `critical' cases the indecomposability of $X$ implies (see \ref{sec:Jordan}) that $H^{nd}$
has degeneracies. One unexpected
bonus is that, using the spectral equivalence, we get also get 
degeneracies in $H^d$ and $H^{lp}$. In \cite{Alcaraz:1988vi,Grimm:1990dg,Grimm:1990gg} it was observed, numerically, that for certain values of $\gamma$ and
$\omega$,
the spectrum of $H^d$ is composed of doublets and singlets. This observation was
used,
with no deep reasons, to obtain the minimal models by disregarding the doublets.
This
ad hoc rule now gets an explanation. The singlets correspond to the `good'
reducible
representations of the 1BTL algebra obtained using the truncated Bratelli diagrams
described in Section \ref{sec:Bratelli} and the doublets correspond to the
Jordan cells in $H^{nd}$. In section 4.5 we showed that for two of the
`critical' cases, namely $\omega= \pm \gamma$, the energy levels of the states
in the truncated space had linear (or no) dependence on the boundary parameter $a$.

It was previously noticed in \cite{Alcaraz:1988zr} that for a mysterious reason, in the continuum
limit, the spectrum of $H^d$ depends only on $\gamma$ and
$\omega$ and not on $\delta$. The origin of the different role played by the three parameters is
now clear. The continuum theory is dependent only on the two parameters
$\gamma$ and $\omega$ which belong to the 1BTL. The equality of the spectra
implies, in particular, that in the finite-size scaling limit, the Coulomb gas
description known for $H^d$ \cite{Saleur:1998hq} is valid also for the other
two Hamiltonians.

If we limit our considerations to the XXZ chain with boundaries, we have seen that
for two disconnected domains in the space of coefficients of the boundary
terms (corresponding to $H^{nd}$ and $H^d$) one
can establish a spectral equivalence. For the decoupling point at
$\Delta=0$ using the results of \cite{BilsteinII} in the continuum
limit one can show that the two domains exhaust the parameter space
 where equivalences to $H^{nd}$ and $H^d$ can be found.

The three representations of the 1BTL algebra
considered here, namely those in $H^{nd}$, $H^d$, and $H^{lp}$, do not exhaust
the possible $2^L$ dimensional representations. Another representation (depending on one free parameter) can be obtained using an ideal of the
two-boundary Temperley-Lieb algebra \cite{deGier:2003iu}. In this representation, similar to the link
representation, one can define a diagrammatic charge and one can control the
existence of Jordan cells through the free parameter. In \cite{Martin:1992td,Martin:1993jk,MartinWoodcockI,MartinWoodcockII} a faithful
representation of 1BTL of dimension $2^L$ was given.

What is next? It is clear that one has to consider the `master' Hamiltonian
with two boundary generators (the two-boundary Temperley-Lieb algebra
\cite{deGier:2003iu}). This will allow us to consider XXZ models with arbitrary boundary terms
at the two ends of the chain and to define more integrable link pattern models. To find
the Jordan cell structures in these cases will be a new exercise. In contrast
to the 1BTL algebra, where the `critical'
values of the parameters are known, almost nothing is known about the 2BTL
algebra.

\renewcommand\thesection{}
\section{Acknowledgements}
A.N. and V.R. are grateful for financial support by the E.U. network {\it Integrable models and
  applications: from strings to condensed matter} HPRN-CT-2002-00325. J. de G. and V.R. gratefully acknowledge the support of the Deutsche Forschungsgemeinschaft and the Australian Research Council. We would like to
  particularly thank F.C. Alcaraz for discussions on the Bethe ansatz 
and to P.N. Pyatov for a detailed reading of the manuscript. We have
also enjoyed stimulating discussions with: A. Belavin, A. Doikou,
P. Martin, and R.I. Nepomechie.
\setcounter{equation}{0} 
%
\appendix \renewcommand\thesection{Appendix \Alph{section}}
\renewcommand{\theequation}{\Alph{section}.\arabic{equation}}
\setcounter{equation}{0} 
\section{Matrices with Jordan cells structure}
\label{sec:Jordan}
In this paper matrices with Jordan cell structures appear in several
different contexts. Here, using simple examples, we would like to
mention two particular properties of such matrices.
 
Let $X$ be a $2$ by $2$ matrix which is of the Jordan form:
\bea X = 
\left (\begin{array}{cc} x & 1 \\ 
0 & x
\end{array}
\right) 
\eea
It is trivial to check that if a matrix $H$ commuting with $X$ has the
form:
\bea \label{eqn:MatrixH} H=\left (\begin{array}{cc} \alpha & \beta \\
0 & \alpha
\end{array}
\right) \eea
then this implies that the matrix $H$ has a degenerate spectrum and
can, but does not have to have, a Jordan form (one can take
$\beta=0$). This observation has an obvious application when $X$ is a
centralizer and $H$ is the Hamiltonian. If $X$ were not of Jordan
form, then the condition that it commutes with $H$ would not give rise
to any degeneracies in the spectrum of the Hamiltonian. However if $X$
is of the Jordan form then the spectrum of $H$ acquires a degeneracy
even though there is only one operator which commutes with it.

We come now to the second observation that a Hamiltonian with Jordan
structure leads to a different time evolution for certain
observables. For example let us assume the Hamiltonian giving the
Euclidean time evolution of a stochastic model \cite{KadanoffSwift}
has the Jordan form given by the matrix $H$ (\ref{eqn:MatrixH}). For
instance one has to compute the quantity
\bea e^{-t H} \left. | P \right>= e^{-\alpha t}\left
(\begin{array}{cc} 1 & -\beta t \\ 0 & 1
\end{array}
\right) \left. | P \right> \eea
where $\left. | P \right>$ is the initial probability
distribution. Notice that the time dependence is \emph{not} purely
exponential. This observation implies in particular that for systems
described by Hamiltonians with Jordan cell structures the finite-size
scaling limit behaviour of various quantities can be very different
than the known ones.
\section{Similarity Transformations for $L=2$ site Hamiltonians}
\label{sec:Similarities}
\setcounter{equation}{0}
The most general similarity transformation is:
\bea U H^{nd}=H^{d} U \eea
which relates the two Hamiltonians $H^{nd}$ (\ref{eqn:Hnd}) and $H^d$
(\ref{eqn:Hd}). For the two site case this is given by:
\bea \label{eqn:similarity} U=\left(
\begin{array}{llll}
e^{i(\gamma + 2 \omega)} x_1 & i e^{i \omega} x_1 & -i
e^{i(\gamma+\omega)} x_1 & x_1 \\ e^{-i\gamma} x_2 &A_{22} &A_{23} &
x_2 \\ e^{-i\gamma} x_3 &A_{32} &A_{33} & x_3 \\ e^{i(\gamma -2
\omega)} x_4 & i e^{-i \omega} x_4 & -i e^{i(\gamma- \omega)} x_4&x_4
\end{array}
\right) \eea
where:
\bea A_{22}&=& - \left\{\left[\cos \delta \cot \gamma + \cos \omega
(-i + \cot \gamma) + \sin \delta \right] x_2 +(\cos \delta + \cos
\omega) \csc \gamma x_3\right\} \nonumber \\ A_{23}&=& -e^{-i \gamma}
\csc \gamma \left[ (\cos(\delta-\gamma)+e^{i \gamma} \cos \omega )x_2
+(\cos \delta +\cos \omega)x_3\right] \\ A_{32}&=& - \left\{\csc
\gamma ( \cos \delta + \cos \omega)x_2+ \left[\cos \omega (-i + \cot
\gamma) +\cos(\delta+\gamma)\csc \gamma \right]x_3\right\} \nonumber\\
A_{33}&=& -e^{-i \gamma} \left\{ \csc \gamma ( \cos \delta +\cos
\omega)x_2 + \left[ \cos \omega (i+\cot \gamma) + \cos(\delta +
\gamma) \csc \gamma \right]x_3 \right\} \nonumber \eea
and $x_i,~(i=1,\cdots,4)$ are free parameters. From
(\ref{eqn:similarity}) we obtain:
\bea Det~U~&=&4 \left[\cos (2 \gamma) -\cos (2 \omega)\right] \csc
\gamma \sin \omega x_1 x_4 \nonumber \\ && \quad \left[ - (\cos \delta
+ \cos \omega )x_2^2+2 \sin \delta \sin \gamma x_2 x_3 +(\cos
\delta+\cos \omega)x_3^2 \right] \eea
One can easily see that if
\bea \label{eqn:exceptionalcases2sites} \omega = -\gamma,0, \gamma
\eea
there is no choice of the parameters $x_i$ such that ${\rm Det~}U$
doesn't vanish. Therefore in these cases the similarity transformation
does not exist.  Notice that in (\ref{eqn:exceptionalcases2sites})
$\delta$ does not appear. This suggests that
(\ref{eqn:exceptionalcases2sites}) is related to the 1BTL algebra and
not specifically to the Hamiltonian. This is indeed the case, since
these points are a special case of (\ref{eqn:boundaryTLcritical})
which defines the `critical' algebras.
 
The reason why the similarity transformation fails for the cases
(\ref{eqn:exceptionalcases2sites}) is the appearance of Jordan cell
structures. The eigenvalues of the two Hamiltonians are the same:
\begin{itemize}
\item{$\omega = -\gamma$}
\bea \lambda_1=-a ~{\rm (twice)}; \quad\lambda_2=0 ; \quad\lambda_3=-2
\cos \gamma \eea
\item{$\omega =0$}
\bea \lambda_1=0~{\rm (twice)}; \quad \lambda_2=-\cos \gamma - \half
\sqrt{2 + 4a + 2 \cos (2 \gamma)} \\ \lambda_3=-\cos \gamma + \half
\sqrt{2 + 4a + 2 \cos (2 \gamma)} \eea
\item{$\omega =\gamma$}
\bea \label{eqn:omega=gammacase} \lambda_1=0~{\rm (twice)};
\quad\lambda_2=-\f{a}{2 \cos \gamma};\quad \lambda_3=-2 \cos
\gamma-\f{a}{2 \cos \gamma} \eea
\end{itemize}
where $a$ is defined in (\ref{eqn:Definitionofa})

In all the exceptional cases (\ref{eqn:exceptionalcases2sites}) we
find that $H^{nd}$ has a Jordan cell structure:
\bea H^{nd} \sim \left(
\begin{array}{llll}
\lambda_1 & 1 & 0 & 0 \\ 0 & \lambda_1 & 0 & 0 \\ 0 & 0 & \lambda_2 &
0 \\ 0 & 0 & 0 & \lambda_3
\end{array}
\right) \eea
while $H^d$ is diagonalizable.

We now turn to the relation between $H^{nd}$ and $H^{lp}$. We mention
only the result. There exits a similarity transformation between the
two Hamiltonians except for the special cases:
\bea \omega = \pi/3, -\gamma.  \eea
It turns out that for these cases $H^{lp}$ is fully diagonalizable
whereas $H^{nd}$ is not. For the remaining cases in
(\ref{eqn:exceptionalcases2sites}) both $H^{lp}$ and $H^{nd}$ are not
fully diagonalizable and so can again be related by a similarity
transformation.
\section{The `diagonal' representation of the one-boundary
Temperley-Lieb algebra}
\label{sec:DiagonalRepn}
\setcounter{equation}{0}
Using the representation given in equations (\ref{eqn:TLgenerators})
and (\ref{eqn:e0}) of the 1BTL algebra we have obtained $H^{nd}$ given
by (\ref{eqn:Hnd}). In Section 2, we mentioned the surprising
observation that the spectrum of $H^{nd}$ coincides with the spectrum
of $H^d$ given in (\ref{eqn:Hd}). The latter Hamiltonian commutes with
the local charge $S^z$ (\ref{eqn:Sz}). A natural question is: Does the
1BTL algebra have a representation in which each generator commutes
with $S^z$?

We consider here the two site problem and look for a $4\times 4$
 representation of the 1BTL algebra with the property:
\bea [S^z, e_0] = 0 = [S^z, e_1] \eea
Such a representation exists:
\bea e_0&=&\left(
\begin{array}{llll}
\f{\sin \omega}{\sin(\omega+\gamma)} & 0 & 0 & 0 \\ 0 & \f{\sin
\omega}{\sin(\omega+\gamma)} & \cos \left(\f{\omega+\delta}{2}\right)
\sec \left(\f{\omega-\delta}{2}\right) \f{\sin
\gamma}{\sin(\omega+\gamma)} & 0 \\ 0 & 0 & 0 & 0 \\ 0 & 0 & 0 & 0
\end{array}
\right) \\ e_1&=&\left(
\begin{array}{llll}
0 & 0 & 0 & 0 \\ 0 & \eta & \eta \xi & 0 \\ 0 & 1 & \xi & 0 \\ 0 & 0 &
0 & 0
\end{array}
\right) \eea
where:
\bea \eta&=&\cos \left( \f{\delta-2 \gamma-\omega}{2} \right) \sec
\left( \f{\delta-\omega}{2} \right) \nonumber \\ \xi&=&\cos \left(
\f{\delta+2 \gamma-\omega}{2} \right) \sec \left( \f{\delta-\omega}{2}
\right) \eea
and $\delta$ is a free parameter. Notice that all parameters
\emph{including} $\delta$ appear in both $e_0$ and $e_1$.
 
Taking $a$ (which depends on $\delta$) with the parameterization given
in (\ref{eqn:Definitionofa}), we observe that the Master Hamiltonian
(\ref{eqn:TLlb}):
\bea H^{M} = -a e_0 - e_1 \eea
coincides with the XXZ Hamiltonian with diagonal boundary terms $H^d$
(\ref{eqn:Hd}) for the two site case.

Let us now perform a similarity transformation:
\bea e_i \rightarrow e_i=U e_i U^{-1} \quad (i = 0, 1) \eea
with:
\bea \label{eqn:Diagsim} U=\left(
\begin{array}{llll}
1 & 0 & 0 & 0 \\ 0 & 1 & \xi - e^{-i \gamma} & 0\\ 0 & 0 & 1 & 0\\ 0 &
0 & 0 & 1
\end{array}
\right) \eea
then we obtain:
\bea \label{eqn:diagonalTL} e_0&=&\left(
\begin{array}{llll}
\f{\sin \omega}{\sin(\omega+\gamma)} & 0 & 0 & 0 \\ 0 & \f{\sin
\omega}{\sin(\omega+\gamma)} & 1- \f{e^{i \gamma} \sin
\omega}{\sin(\omega+\gamma)} & 0 \\ 0 & 0 & 0 & 0 \\ 0 & 0 & 0 & 0
\end{array}
\right) \\ e_1&=&\left(
\begin{array}{llll}
0 & 0 & 0 & 0 \\ 0 & e^{i \gamma} & 1 & 0 \\ 0 & 1 & e^{- i \gamma} &
0 \\ 0 & 0 & 0 & 0
\end{array}
\right) \eea
The similarity transformation (\ref{eqn:Diagsim}) leaves $S^z$
unchanged however it brings $e_1$ to the standard form
(\ref{eqn:TLgenerators}) and removes the parameter $\delta$ from
$e_0$. The crucial difference between this `diagonal' representation
(\ref{eqn:diagonalTL}) and the `non-diagonal' representation
(\ref{eqn:e0}) of 1BTL is that $e_0$ acts now in the spin chain not
only on site $1$ but on the two sites $1$ \emph{and} $2$. The
representation of the 1BTL given in (\ref{eqn:diagonalTL}) generalizes
\cite{WorkInProgress} - the bulk generators $e_i$ retain their
standard form (\ref{eqn:TLgenerators}) and the generator $e_0$ now
acts on \emph{all} the $L$ sites of the spin chain and commutes with
$S^z$.

It is important to stress that the `non-diagonal' and the `diagonal'
representations are equivalent except for the cases
(\ref{eqn:boundaryTLcritical}) when the 1BTL algebra is `critical'. In
these `critical' cases the `non-diagonal' representation stays
faithful whereas the `diagonal' one is not (see
\ref{sec:Faithfulness}).
\section{Faithfulness of the representations}
\label{sec:Faithfulness}
\setcounter{equation}{0}
In this paper we have used several different representations of the
1BTL algebra. Although these are all of dimension $2^L$ we have seen
that there can be a difference in the Jordan cell structures which
appear. In this section we shall show that the question of possible
Jordan cell structure in the Hamiltonians is intimately linked to
their faithfulness. Here we shall only comment on the example of two
sites.

At two sites using the 1BTL algebra we can construct only $6$ words:
\bea {\bf 1}, ~e_0, ~e_1, ~e_1 e_0, ~e_0 e_1, ~e_0 e_1 e_0 \eea
If a representation is faithful for all values of the parameters
$\gamma$ and $\omega$ then there should \emph{not} exist any linear
relations between these words.

In the non-diagonal representation at all values of the parameters
$\gamma$ and $\omega$ we did not find any linear relations between the
words and therefore we conclude that at $L=2$ sites this
representation is indeed faithful. It is possible that this is true in
general but we have no proof.

In the diagonal and link pattern representations we again found that
for generic values of the parameters $\gamma$ and $\omega$ these were
faithful. However for particular cases there were additional kernals
in the representations. (For the case $\omega=-\gamma$ we rescale
$e_0$ and have $e_0^2=e_0$ and $e_1 e_0 e_1=0$.)

In the diagonal case the kernals are given by:
\begin{itemize}
\item{$\omega=0$}
\bea e_0 - e_0 e_1 e_0 =0 \eea
\item{$\omega=-\gamma$}
\bea e_0 e_1 =0 \eea
\item{$w=\gamma$}
\bea \label{eqn:DiagonalKernal} 2 \cos \gamma e_1 e_0 - e_1=0 \eea
\end{itemize}
In the link-pattern case the kernals are given by:
\begin{itemize}
\item{$\gamma=\f{\pi}{3},~\omega=\f{\pi}{3}$}
\bea \label{eqn:LeftIdealquotient} e_1 e_0 -e_1=0 \eea
\item{$\omega=-\gamma$}
\bea e_0 e_1=0 \eea
\end{itemize}
These kernals are responsible for the disappearance of Jordan cell
structure from the `master' Hamiltonian (\ref{eqn:TLlb}). Let us
illustrate this with an example $\omega=\gamma$. Using the 1BTL we
find that the following is identically satisfied:
\bea \label{eqn:fulleqn} \left(H^M - \lambda_1 \right)^2 \left(H^M -
\lambda_2 \right) \left(H^M - \lambda_3 \right)=0 \eea
where the $\lambda_i$ were given in (\ref{eqn:omega=gammacase}). This
is equivalent to the statement that the `master' Hamiltonian can be
brought into the form:
\bea \label{Hmdiag} H^M \sim \left(
\begin{array}{llll}
\lambda_1 & c & 0 & 0 \\ 0 & \lambda_1 & 0 & 0 \\ 0 & 0 & \lambda_2 &
0 \\ 0 & 0 & 0 & \lambda_3
\end{array}
\right) \eea
where $c$ can take any value.

Now let us consider the case in which we can simplify
(\ref{eqn:fulleqn}) to:
\bea \label{eqn:reducedeqn} \left(H^M - \lambda_1 \right) \left(H^M -
\lambda_2 \right) \left(H^M - \lambda_3 \right)=0 \eea
This is equivalent to the statement that $c=0$ in (\ref{Hmdiag})
i.e. the `master' Hamiltonian (\ref{eqn:TLlb}) has no Jordan cell
structure and can be completely diagonalized. Using just the
one-boundary \TL algebra we find (\ref{eqn:reducedeqn}) is \emph{not}
an identity. Hence if the representation is faithful then the `master'
Hamiltonian cannot be fully diagonalized - as indeed observed for
$H^{nd}$.

In the diagonal representation the L.H.S. of (\ref{eqn:reducedeqn}) is
 proportional to the kernal (\ref{eqn:DiagonalKernal}) and therefore
 does vanish. This implies that, in the diagonal representation, there
 is no Jordan cell structure in the `master' Hamiltonian - as can be
 directly verified. All other differences 
 of Jordan cell structure of Hamiltonians can
be understood in a similar way.
\section{The spectra of $H^{nd}$, $H^d$ and $H^{lp}$: The Bethe ansatz}
\label{sec:Betheansatz}
\setcounter{equation}{0}
In \cite{deGier:2003iu} the loop Hamiltonian $H^{lp}$ was investigated and, as shown in
Section $3$, it has a block triangular form. In order to compute the
spectrum it is sufficient to consider the block diagonal part which
conserves the diagrammatic charge $C$. Any indecomposable structure
that occurs between these block diagonal parts is simply ignored. One
can write Bethe ansatz equations for this 
block diagonal part considering separately the case $C$
even and $C$ odd (see Tables $1$ and $2$ in Section $3$). The reference states
have $C=0$ and $C=1$ respectively. In \cite{deGier:2003iu} it was shown that the
Bethe ansatz equations coincide with those obtained or conjectured in
\cite{ChineseGuys,Nepomechie:2003ez} for the Hamiltonian
$H^{nd}$. What remains to be shown is that the Hamiltonian 
$H^d$ (\ref{eqn:Hd}) has the same spectrum as $H^{lp}$.

We start (exactly as in \cite{Alcaraz:1987uk}) with the Bethe
reference state with all spins up and then add $M$ down spins. We find
the energy eigenvalues are given by: 
\bea
E=-\sum_{i=1}^M \left(t+z_i + z_i^{-1}\right)
\eea
where $t=2 \cos \gamma$ and the $z_i$ are solutions of:
\bea
z_i^{2L}=\f{K(z_i)}{K(z_i^{-1})}\prod_{j=1}^{M}\f{S(z_i^{-1},z_j) S(z_j,z_i)}{S(z_j,z_i^{-1}) S(z_i,z_j)}
\eea
with:
\bea
K(z)=t+z+z^{-1}-a(s+z) \quad \quad S(z,w)=1+t w + z w
\eea
and $s=\f{\sin \omega}{\sin(\omega+\gamma)}$. If $M \le \f{L}{2}$ then this
result exactly agrees with the eigenvalues of the loop Hamiltonian with $C$
even. 
However one may also start in the spin chain from the reference state
with all spins down and add $M$ spin up waves. Then we find that in
the sector with $M$ spins up: 
\bea
E=-as-\sum_{i=1}^M \left(t+z_i + z_i^{-1}\right)
\eea
with the $z_i$ given by:
\bea
z_i^{2L}=\f{K(z_i)}{K(z_i^{-1})}\prod_{j=1}^{M}\f{S(z_i^{-1},z_j) S(z_j,z_i)}{S(z_j,z_i^{-1}) S(z_i,z_j)}
\eea
where:
\bea
K(z)=t+z+z^{-1}+a(s+z(st-1)) \quad \quad S(z,w)=1+t w + z w
\eea
This result now agrees for $M\leq L/2$ with the eigenvalues of the
loop Hamiltonian with $C$ odd. The first set of equations with $M>L/2$ does not give the same
solutions as the second set with $M^*=L-M$. In fact, preliminary
numerical studies indicate that the first set produces incorrect
eigenvalues for such values of $M$. The complete spectrum is described
by using both sets of equations and restricting $M\leq L/2$. We refer to 
\cite{Nepomechie:2003ez} for similar numerical observations and 
\cite{Caux:2003mw} for related analytic results in the continuum
limit.  

The Bethe ansatz equations for the XXZ model with periodic boundary
conditions has been investigated in detail in
\cite{Baxter:2001sx}. Again we can start from two different reference
states: one with all spins up and the other with all spins down. Here
to be `on the wrong side of the equator', i.e. $M > \f{L}{2}$,
does not give rise to unresolvable problems and the Bethe ansatz
solutions, including those `over the equator', give the complete set of
eigenstates.   

The problem of being `on the wrong side of the equator' can be seen most
vividly in the case $st=1$. Then one set of equations contains a
manifest $a$ dependence whereas the other does not. This case is
interesting due to its relevance for stochastic processes at $t=1,s=1$
\cite{RefinedJanandVladimir}. The complete set of wavefunctions is
only obtained by combining those obtained from \emph{both} Bethe
reference states.  


\begin{thebibliography}{10}

\bibitem{Essler}
F.~Essler and R.~Konik,
\newblock Application of massive integrable QFTs to problems in condensed
  matter physics,
\newblock in {\em From fields to Strings: Circumnavigating Theoretical Physics,
  Ian Kogan Memorial Volume}, edited by M.~Shifman, A.~Vainshtein, and
  J.~Wheater Vol.~2, http://www.hep.umn.edu/\~{}vainshte/Kogan/

\bibitem{RaiseandPeel}
J.~de~Gier, B.~Nienhuis, P.~A. Pearce, and V.~Rittenberg,
\newblock J. Statist. Phys. {\bf 114}, 1 (2004), cond-mat/0301430.

\bibitem{PyatovHexagon}
P.~Pyatov,
\newblock J. Stat. Mech.: Theor. Exp. , P09003 (2004), math-ph/0406025.

\bibitem{Razumov:2000ei}
A.~V. Razumov and Y.~G. Stroganov,
\newblock J. Phys. {\bf A34}, 3185 (2001), cond-mat/0012141.

\bibitem{JanReview}
J.~de~Gier,
\newblock Loops, matchings and alternating-sign matrices,
\newblock in {\em 14th International Conference on Formal Power Series and
  Algebraic Combinatorics (Melbourne 2002)}, math.CO/0211285.

\bibitem{Alcaraz:1987uk}
F.~C. Alcaraz, M.~N. Barber, M.~T. Batchelor, R.~J. Baxter, and G.~R.~W.
  Quispel,
\newblock J. Phys. {\bf A20}, 6397 (1987).

\bibitem{Alcaraz:1988zr}
F.~C. Alcaraz, M.~N. Barber, and M.~T. Batchelor,
\newblock Ann. Phys. {\bf 182}, 280 (1988).

\bibitem{Pasquier:1989kd}
V.~Pasquier and H.~Saleur,
\newblock Nucl. Phys. {\bf B330}, 523 (1990).

\bibitem{TemperleyLieb}
H.~N.~V. Temperley and E.~Lieb,
\newblock Proc. Roy. Soc. London {\bf A322}, 251 (1971).

\bibitem{MartinBook}
P.~P. Martin,
\newblock {\em Potts Models and Related Problems in Statistical Mechanics}
  (World Scientific, 1991).

\bibitem{Alcaraz:1987ni}
F.~C. Alcaraz, M.~Baake, U.~Grimm, and V.~Rittenberg,
\newblock J. Phys. {\bf A21}, L117 (1988).

\bibitem{Alcaraz:1988vi}
F.~C. Alcaraz, M.~Baake, U.~Grimm, and V.~Rittenberg,
\newblock J. Phys. {\bf A22}, L5 (1989).

\bibitem{Grimm:1990dg}
U.~Grimm and V.~Rittenberg,
\newblock Int. J. Mod. Phys. {\bf B4}, 969 (1990), hep-th/0311085.

\bibitem{Grimm:1990gg}
U.~Grimm and V.~Rittenberg,
\newblock Nucl. Phys. {\bf B354}, 418 (1991).

\bibitem{deVega:1992zd}
H.~J. de~Vega and A.~Gonzalez~Ruiz,
\newblock J. Phys. {\bf A26}, L519 (1993), hep-th/9211114.

\bibitem{BilsteinI}
U.~Bilstein and B.~Wehefritz,
\newblock J. Phys. {\bf A 32}, 191 (1999), cond-mat/9807166.

\bibitem{ChineseGuys}
J.~Cao, H.-Q. Lin, K.-J. Shi, and Y.~Wang,
\newblock Nucl. Phys. {\bf B663}, 487 (2003), cond-mat/0212163.

\bibitem{Nepomechie:2002xy}
R.~I. Nepomechie,
\newblock J. Statist. Phys. {\bf 111}, 1363 (2003), hep-th/0211001.

\bibitem{Nepomechie:2003vv}
R.~I. Nepomechie,
\newblock J. Phys. {\bf A37}, 433 (2004), hep-th/0304092.

\bibitem{Nepomechie:2003ez}
R.~I. Nepomechie and F.~Ravanini,
\newblock J. Phys. {\bf A36}, 11391 (2003), hep-th/0307095.

\bibitem{deGier:2003iu}
J.~de~Gier and P.~Pyatov,
\newblock J. Stat. Mech.: Theor. Exp. , P03002 (2004), hep-th/0312235.

\bibitem{Martin:1992td}
P.~Martin and H.~Saleur,
\newblock Commun. Math. Phys. {\bf 158}, 155 (1993), hep-th/9208061.

\bibitem{Martin:1993jk}
P.~Martin and H.~Saleur,
\newblock Lett. Math. Phys. {\bf 30}, 189 (1994), hep-th/9302094.

\bibitem{MartinWoodcockI}
P.~P. Martin and D.~Woodcock,
\newblock Journal of Algebra {\bf 225}, 957 (2000).

\bibitem{MartinWoodcockII}
P.~P. Martin and D.~Woodcock,
\newblock LMS J. Comput. Math. {\bf 6}, 249 (2003), math.RT/0205263.

\bibitem{Delius:2001qh}
G.~W. Delius and N.~J. MacKay,
\newblock Commun. Math. Phys. {\bf 233}, 173 (2003), hep-th/0112023.

\bibitem{Delius:2002mv}
G.~W. Delius and A.~George,
\newblock Quantum group symmetry of integrable models on the half- line,
\newblock in {\em Workshop On Integrable Theories, Solitons And Duality (Sao
  Paulo, Brazil 2002)}, JHEP, 2002, hep-th/0212300.

\bibitem{Mezincescu:1997nw}
L.~Mezincescu and R.~I. Nepomechie,
\newblock Int. J. Mod. Phys. {\bf A13}, 2747 (1998), hep-th/9709078.

\bibitem{Doikou:2004km}
A.~Doikou,
\newblock (2004), math-ph/0402067.

\bibitem{WorkInProgress}
A.~Nichols,
\newblock In preparation.

\bibitem{Baxter:2001sx}
R.~J. Baxter,
\newblock J. Stat. Phys. {\bf 108}, 1 (2002), cond-mat/0111188.

\bibitem{RefinedJanandVladimir}
J.~de~Gier and V.~Rittenberg,
\newblock J. Stat. Mech.: Theor. Exp. , P09009 (2004), math-ph/0408042.

\bibitem{Belavin:2003id}
A.~A. Belavin,
\newblock J. Phys. {\bf A37}, 317 (2004), hep-th/0305209.

\bibitem{Doikou:2002ry}
A.~Doikou and P.~P. Martin,
\newblock J. Phys. {\bf A36}, 2203 (2003), hep-th/0206076.

\bibitem{Saleur:1998hq}
H.~Saleur,
\newblock Lectures on non perturbative field theory and quantum impurity
  problems,
\newblock in {\em Topological aspects of low dimensional systems (Les Houches
  1998)}, edited by A.~Comtet, T.~Jolicoeur, S.~Ouvry, and F.~David, Springer,
  1999, cond-mat/9812110.

\bibitem{BilsteinII}
U.~Bilstein,
\newblock J. Phys. {\bf A 33}, 4437 (2000), cond-mat/0002162.

\bibitem{KadanoffSwift}
L.~P. Kadanoff and J.~Swift,
\newblock Phys. Rev. {\bf 165}, 310 (1968).

\bibitem{Caux:2003mw}
J.~S. Caux, H.~Saleur, and F.~Siano,
\newblock Nucl. Phys. {\bf B672}, 411 (2003), cond-mat/0306328.

\end{thebibliography}

\end{document}